%% file: bare_jrnl.tex
\documentclass[journal, onecolumn]{IEEEtran}
\ifCLASSINFOpdf
\else
\fi
\usepackage[moderate]{savetrees}
\usepackage{xspace}
\usepackage[hyphens]{url}
\usepackage{enumerate}
\usepackage[shortlabels]{enumitem}
\usepackage{amssymb}
\usepackage{amsmath}
\allowdisplaybreaks 
\usepackage{amsthm} 
\usepackage{graphicx}
\usepackage{color}
\usepackage{alltt}
\usepackage{algorithm}
\usepackage{algorithmic}
\usepackage{setspace}
\usepackage{subcaption}
\usepackage{tabularx}
\usepackage[T1]{fontenc}
\usepackage[utf8]{inputenc}
\DeclareCaptionType{copyrightbox}

\newcommand{\sys}{Drynx\xspace}
\newcommand{\robust}{\textit{robustness}\xspace}
\newcommand{\sbline}{\\[.5\normalbaselineskip]}% small blank line
\newcommand{\enco}{\textit{encoding}\xspace}
\newcommand{\nenco}{\textit{neutral response}\xspace}

\newif\ifcomment
\commentfalse

\ifcomment
	\newcommand{\david}[1]{\textcolor{blue}{#1}}
\else  
    \newcommand{\david}[1]{\textcolor{black}{#1}}
\fi

\newtheorem{definition}{Definition}

\newcounter{protocol}
\newenvironment{protocol}[1]
{\par\addvspace{\topsep}
	\noindent
	\tabularx{\linewidth}{@{} X @{}}
	\hline
	\refstepcounter{protocol}\textbf{Protocol \theprotocol} #1 \\
	\hline
}
{ \\
	\hline
	\endtabularx
	\par\addvspace{\topsep}}

\begin{document}
%
% paper title
% Titles are generally capitalized except for words such as a, an, and, as,
% at, but, by, for, in, nor, of, on, or, the, to and up, which are usually
% not capitalized unless they are the first or last word of the title.
% Linebreaks \\ can be used within to get better formatting as desired.
% Do not put math or special symbols in the title.
\title{\sys: Decentralized, Secure, Verifiable System for Statistical Queries and Machine Learning on Distributed Datasets}
%
%
% author names and IEEE memberships
% note positions of commas and nonbreaking spaces ( ~ ) LaTeX will not break
% a structure at a ~ so this keeps an author's name from being broken across
% two lines.
% use \thanks{} to gain access to the first footnote area
% a separate \thanks must be used for each paragraph as LaTeX2e's \thanks
% was not built to handle multiple paragraphs
%

\author{David~Froelicher,
	Juan R.~Troncoso-Pastoriza\IEEEmembership{, Senior Member, IEEE},
	Joao~Sa Sousa,
	and~Jean-Pierre~Hubaux\IEEEmembership{, Fellow, IEEE}% <-this % stops a space
	\thanks{This work was partially supported by the grant \#2017-201 of the Strategic Focal Area “Personalized Health and Related Technologies (PHRT)” of the ETH Domain.}
	\thanks{D. Froelicher is with the Laboratory for Data Security and DeDiS Laboratory, Ecole Polytechnique Federale de Lausanne, 1015 Lausanne, Switzerland, e-mail: david.froelicher@epfl.ch. % <-this % stops a space
	Joao Sa Sousa, Juan R. Troncoso-Pastoriza and Jean-Pierre~Hubaux are with the Laboratory for Data Security, Ecole Polytechnique Federale de Lausanne, 1015 Lausanne, Switzerland, e-mail: name.surname@epfl.ch.}% <-this % stops a space
	%\thanks{Manuscript received X, 2019; revised X, XXXX.}}
	}

\maketitle

% As a general rule, do not put math, special symbols or citations
% in the abstract or keywords.
\input{abs}

% Note that keywords are not normally used for peerreview papers.

\input{introduction}
\input{relatedWork}
\input{background}
\input{model}
\input{design}
\input{securityAnalysis}
\input{encoding}

\input{discussion}
\input{evaluation}
\input{conclusion}

% For peer review papers, you can put extra information on the cover
% page as needed:
% \ifCLASSOPTIONpeerreview
% \begin{center} \bfseries EDICS Category: 3-BBND \end{center}
% \fi
%
% For peerreview papers, this IEEEtran command inserts a page break and
% creates the second title. It will be ignored for other modes.
\IEEEpeerreviewmaketitle

\section*{Acknowledgment}
The authors would like to thank Henry Corrigan-Gibbs and all members of the Laboratory for Data Security at EPFL for their helpful feedback and their support.

% Can use something like this to put references on a page
% by themselves when using endfloat and the captionsoff option.
\ifCLASSOPTIONcaptionsoff
  \newpage
\fi

% trigger a \newpage just before the given reference
% number - used to balance the columns on the last page
% adjust value as needed - may need to be readjusted if
% the document is modified later
%\IEEEtriggeratref{8}
% The "triggered" command can be changed if desired:
%\IEEEtriggercmd{\enlargethispage{-5in}}

% references section

% can use a bibliography generated by BibTeX as a .bbl file
% BibTeX documentation can be easily obtained at:
% http://mirror.ctan.org/biblio/bibtex/contrib/doc/
% The IEEEtran BibTeX style support page is at:
% http://www.michaelshell.org/tex/ieeetran/bibtex/
\bibliographystyle{IEEEtran}
% argument is your BibTeX string definitions and bibliography database(s)
\bibliography{vldb_sample}

\input{appendix}
%
% <OR> manually copy in the resultant .bbl file
% set second argument of \begin to the number of references
% (used to reserve space for the reference number labels box)
%\begin{thebibliography}{1}

%\bibitem{IEEEhowto:kopka}
%H.~Kopka and P.~W. Daly, \emph{A Guide to \LaTeX}, 3rd~ed.\hskip 1em plus
%  0.5em minus 0.4em\relax Harlow, England: Addison-Wesley, 1999.

%\end{thebibliography}

% biography section
% 
% If you have an EPS/PDF photo (graphicx package needed) extra braces are
% needed around the contents of the optional argument to biography to prevent
% the LaTeX parser from getting confused when it sees the complicated
% \includegraphics command within an optional argument. (You could create
% your own custom macro containing the \includegraphics command to make things
% simpler here.)
%\begin{IEEEbiography}[{\includegraphics[width=1in,height=1.25in,clip,keepaspectratio]{mshell}}]{Michael Shell}
% or if you just want to reserve a space for a photo:

\newpage
\begin{IEEEbiography}[{\includegraphics[width=1in,height=1.25in,clip,keepaspectratio]{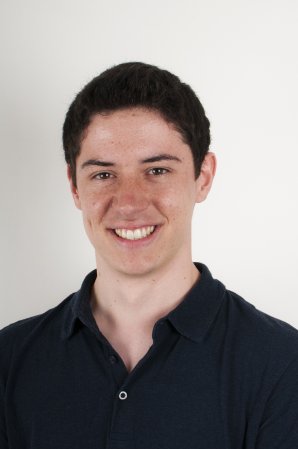}}]{David Froelicher}
	is a PhD candidate at EPFL under the direction of professors Jean-Pierre Hubaux and Bryan Ford. He earned his MS and BS degree in Communication Systems and IT Security at EPFL and did a 6-month internship at NEC. His main research interests are in applied cryptography, privacy and decentralized systems. 
\end{IEEEbiography}

\begin{IEEEbiography}[{\includegraphics[width=1in,height=1.25in,clip,keepaspectratio]{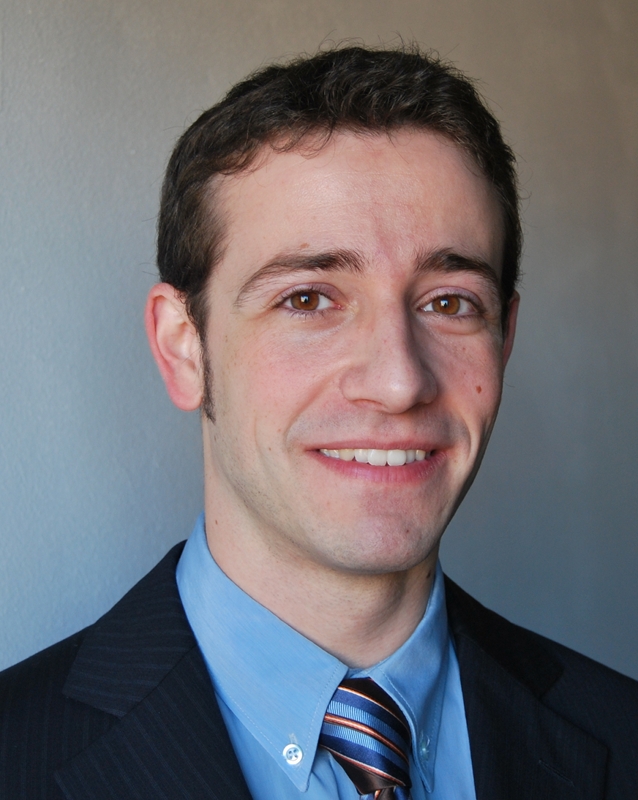}}]{Juan Ram\'on Troncoso-Pastoriza,}
	IEEE Senior Member, senior researcher at the Laboratory of Data Security, EPFL, Switzerland. His research work is focused on applied cryptography for the protection of sensitive signals in distributed and outsourced environments, with a special interest in medical environments and genomic privacy. He actively contributes to the Homomorphic Encryption standardization efforts (\url{https://homomorphicencryption.org}), and to the design and development of the Lattigo library (\url{https://github.com/ldsec/lattigo}). He has coauthored numerous works and holds five international patents in the field of information security, and has been part of the organizing committee and TPC of more than 20 workshops and conferences in this area. He has been the scientific coordinator of the H2020 project WITDOM, and is currently an associate editor of four journals on Information Security.
\end{IEEEbiography}

\begin{IEEEbiography}[{\includegraphics[width=1in,height=1.25in,clip,keepaspectratio]{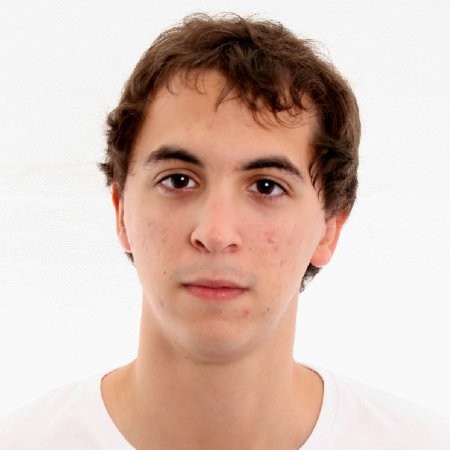}}]{Joao Sa Sousa}
	is currently a Security / Privacy Software Engineer at EPFL under the direction of professor Jean-Pierre Hubaux. He has a MS and BS degree in Informatics Engineering at the University of Coimbra and did a 3-month internship at CMU-SV. His main interests include Wireless Security, Genomic Privacy, Cryptography, Android Development, Web Development and Business Management.
\end{IEEEbiography}

\begin{IEEEbiography}[{\includegraphics[width=1in,height=1.25in,clip,keepaspectratio]{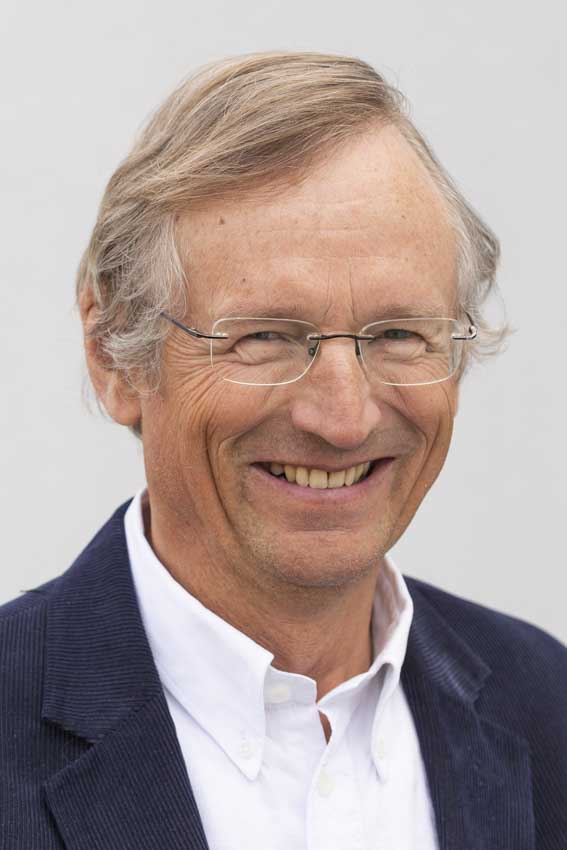}}]{Jean-Pierre Hubaux,}
	IEEE Fellow, is a full professor in the School of Information and Communication Sciences at EPFL and head of the Laboratory for Data Security. Through his research, he contributes to laying the foundations and developing the tools for protecting privacy in today’s hyper-connected world. He has pioneered the areas of privacy and security in mobile/wireless networks and in personalized health. 
	
	He is the academic director of the Center for Digital Trust (C4DT). He leads the  Data Protection in Personalized Health (DPPH) project funded by the ETH Council and is a co-chair of the Data Security Work Stream of the Global Alliance for Genomics and Health (GA4GH). He is a Fellow of both IEEE (2008) and ACM (2010). Recent awards: two of his papers obtained distinctions at the IEEE Symposium on Security and Privacy in 2015 and 2018. 
\end{IEEEbiography}
% insert where needed to balance the two columns on the last page with
% biographies
%\newpage

% You can push biographies down or up by placing
% a \vfill before or after them. The appropriate
% use of \vfill depends on what kind of text is
% on the last page and whether or not the columns
% are being equalized.

%\vfill

% Can be used to pull up biographies so that the bottom of the last one
% is flush with the other column.
%\enlargethispage{-5in}

% that's all folks
\end{document}

%% file: abs.tex
\begin{abstract}
Data sharing has become of primary importance in many domains such as big-data analytics, economics and medical research, but remains difficult to achieve when the data are sensitive. In fact, sharing personal information requires individuals' unconditional consent or is often simply forbidden for privacy and security reasons. In this paper, we propose \sys, a decentralized system for privacy-conscious statistical analysis on distributed datasets. \sys relies on a set of computing nodes to enable the computation of statistics such as standard deviation or extrema, and the training and evaluation of machine-learning models on sensitive and distributed data. To ensure data confidentiality and the privacy of the data providers, \sys combines interactive protocols, homomorphic encryption, zero-knowledge proofs of correctness, and differential privacy. It enables an efficient \david{and decentralized verification of the input data and of all the system's computations
%by relying on a public immutable distributed ledger. It
thus provides auditability in a strong adversarial model in which no entity has to be individually trusted.} \sys is highly modular, dynamic and parallelizable. Our evaluation shows that it enables the training of a logistic regression model on a dataset (\david{12 features and 600,000 records) distributed among 12 data providers in less than 2 seconds. The computations are distributed among 6 computing nodes, and \sys enables the verification of the query execution's correctness in less than 22 seconds}.	 
\end{abstract}

%% file: introduction.tex
\section{Introduction}\label{intro}
To produce meaningful results, statistical and machine-learning analyses often demand large amounts of data. Although data storage and computation costs have dropped over the years, notably due to low-cost and powerful cloud-computing solutions, the sharing of these data is still cumbersome. Massive amounts of data are generated daily to track individuals’ actions, health, shopping habits, interests, political and religious views \cite{techrepublic}, but privacy concerns and ethical/legal constraints often prohibit or discourage the sharing of personal and sensitive data. In Europe, the new data-protection regulation, General Data Protection Regulation (GDPR) \cite{gdpr}, effective since May 2018, requires that (a) the collection and use of personal data can only be done with the consent of the subject and (b) that the data have to be anonymized or encrypted before being shared. This leads to a conundrum, especially in domains such as demography, finance and health, where data have to be shared, e.g., for enabling research, but they also need to be protected to ensure individuals' fundamental right to privacy. Cross-border data sharing is even more challenging, as the legislations among countries can be heterogeneous, forcing companies to geographically adapt their own privacy measures.

Multiple examples show that even when data can be shared, a centralization of the data can have serious consequences, affecting hundreds of millions of individuals \cite{breach2, equifax}; this was the case with the Equifax breach \cite{equifax}, in which personal information (including social-security numbers and credit-card information) of more than 143 million consumers (about 40\% of the US population) was compromised. Centralized solutions are subject to multiple threats as the central database, which stores data from multiple mutually-untrusted sources, constitutes a high-value target for possible attackers and a single point of failure.

Existing solutions for secure databases \cite{bindschaedler2017plausible, hu2015differential, johnson2018towards, popa2011cryptdb, tu2013processing} usually add a cryptographic layer on top of the query engine or focus exclusively on the data-release privacy, e.g., by using differential privacy. However, most of these solutions have a significant performance overhead or are still fully centralized hence either have a single point of failure, or do not protect the data during the query execution.

In this context, decentralized data-sharing systems \cite{bater2017smcql, chen2012towards, jagadeesh2017deriving, kim2018secure, melis2015efficient, raisaro2018medco} have raised considerable interest and are key enablers for privacy-conscious big-data analysis. By distributing the storage and the computation, thus avoiding single points of failure, these systems enable data sharing and minimize the risks incurred by centralized solutions. Nevertheless, many of these systems rely on honest-but-curious or trusted third-party assumptions that might not provide sufficient guarantees when the data to be shared are highly sensitive, valuable, influential or private. Other solutions with stronger threat models, e.g., UnLynx \cite{froelicher2017unlynx},  are limited in the computations they support, e.g., sum only. Moreover, none of these solutions considers the possibility that both computing entities and data providers can be malicious.

Improving upon and using some techniques introduced in UnLynx, we propose \sys, an operational, decentralized and secure system that enables queriers to compute statistical functions and to train and evaluate machine-learning models on data hosted at different sources, i.e., on distributed datasets. \sys ensures data confidentiality, data providers' ($DPs$) privacy and protects individuals' data from potential inferences stemming from the release of end results, i.e., it ensures differential privacy. It also provides computation correctness. Finally, it ensures that strong outliers, either maliciously or erroneously input by $DPs$, cannot influence the results beyond a certain limit, and we denote this by \textit{results} \robust. These guarantees are ensured in a strong adversarial model where no entity has to be individually trusted and a fraction of the system's entities can be malicious. \sys relies on interactive protocols, homomorphic encryption, zero-knowledge proofs of correctness and distributed differential privacy. It is scalable, dynamic and modular: Any entity can leave or join the system at any time and \sys offers security features or properties that can be enforced depending on the application, e.g., differential privacy.
\sbline
In this paper, we make the following contributions:
%enable several databases, as federated that enables privacy-preserving and we abstract the query-engine
\begin{itemize}[leftmargin=*]
	\item We propose \sys, an efficient, modular and parallel system that enables privacy-preserving statistical queries and the training and evaluation of machine-learning regression models on distributed datasets.
	\item We present a system that provides data confidentiality and individuals' privacy, even in the presence of a strong adversary. It ensures the correctness of the computations, protects data providers' privacy and guarantees robustness of query results. 
	\item We propose techniques that enable full and lightweight auditability of query execution. \sys relies on a new efficient distributed solution for storing and verifying proofs of query validity, computation correctness, and input data ranges. \david{We exemplify and evalutate the implementation of this solution by using a blockchain.}
	\item We propose and implement an efficient, modular and multi-functionality query-execution pipeline by
	\begin{itemize}[leftmargin=*]
		\item introducing \textit{Collective Tree Obfuscation}, a new distributed protocol that enables a collective and verifiable obfuscation of encrypted data;
		\item presenting multiple data-encoding techniques that enable distributed computations of advanced statistics on homomorphically encrypted data. We propose new encodings, and improvements and adaptations of previously introduced private-aggregation encodings to our framework and security model;
		\item adapting an existing zero-knowledge scheme for input-range validation to our security model;
		\item proposing a new construction of the \textit{Key Switching} protocol introduced in UnLynx \cite{froelicher2017unlynx}, improving both its performance and capabilities.
	\end{itemize}
\end{itemize}
To the best of our knowledge, \sys is the only operational system that provides the aforementioned security and privacy guarantees. \sys implementation is fully available at \url{www.github.com/ldsec/drynx}.

%% file: relatedWork.tex
%\vspace{-0.5em}
\section{Related Work}\label{sec:related}
Centralized systems for privacy-preserving data sharing \cite{popa2011cryptdb, peer, dong2014achieving, liu2013mona} and trusted-hardware based solutions \cite{ohrimenko2016oblivious} usually require one entity, i.e., a central entity or a hardware provider, to be trusted, which constitutes a single point of failure. Even though these systems can be more efficient than their decentralized counterparts, they often require a centralization or outsourcing of the data storage, which goes against regulations or is cumbersome to achieve \cite{bogdanov2016students} and can be inappropriate for sensitive data. In \sys, we avoid these issues by decentralizing data-storage, computation and correctness verification, thus efficiently distributing trust.

In order to execute queries and compute statistics on distributed datasets, multiple decentralized solutions  \cite{bater2017smcql, jagadeesh2017deriving, melis2015efficient, bogdanov2008sharemind, gascon2017privacy, shokri2015privacy, yang2018private} rely on techniques that have a high expressive power, such as secret sharing and garbled circuits. These solutions are often flexible in the computations they offer but usually assume (a) honest-but-curious computing parties and (b) no collusion or a 2-party model. Furthermore, they do not provide a way to check the computations undertaken in the system. Although they might efficiently distribute trust, their strong honesty assumptions are risky when the data or the computed statistics are highly sensitive. Bater et al. \cite{bater2017smcql} enable the evaluation of various SQL queries on datasets hosted by a set of distrustful data providers, but both the data providers and the computing entity are trusted to follow the protocol. Corrigan-Gibbs and Boneh \cite{corrigan2017prio} propose Prio, a system that ensures privacy as long as one computing entity out of $n$ is honest, but it only guarantees end results robustness in the case where the involved parties are all honest-but-curious. \david{Moreover, Prio does not protect against DPs colluding among themselves or with the computing nodes.} In \sys, no entity has to be individually trusted in order to provide both privacy and robustness.

Systems relying on homomorphic encryption \cite{chen2012towards, kim2018secure, froelicher2017unlynx, nikolaenko2013privacy, papadimitriou2016big, du2018privacy} are often limited in the functionalities they offer (e.g., sum only). They present high-performance overhead in comparison with their less secure counterparts or still rely on honest-but-curious parties. In our previous work, we presented UnLynx \cite{froelicher2017unlynx}, a decentralized system that enables the computation of (only) sums on distributed datasets and ensures $DPs$' privacy and data confidentiality. UnLynx assumes $DPs$ to be honest-but-curious and, unlike \sys, it does not ensure end results robustness. Moreover, UnLynx does not provide a practical solution for auditability. In this work, we show how to overcome these limitations and provide a system that enables secure computations of multiple operations in a stronger threat model.

There are multiple solutions proposed for the problem of training machine-learning models on distributed data in a privacy-preserving way \cite{kim2018secure, nikolaenko2013privacy, mohassel2017secureml, bourahigh, juvekar2018gazelle, aono2016scalable, aono2018privacy, jia2018preserving, popa2019helen}. Mohassel and Zhang \cite{mohassel2017secureml} propose a two-party solution, SecureML; it enables the training of specific models, e.g., linear regression. Boura et al. \cite{bourahigh} present a solution that relies on a novel and more flexible approximation of the logistic regression function but assumes honest-but-curious parties.  Nikolaenko et al. \cite{nikolaenko2013privacy} and Juvekar et al. \cite{juvekar2018gazelle} combine homomorphic encryption and garbled circuits to perform private ridge-regression and neural-network inference, respectively. Aono et al. \cite{aono2016scalable} and Kim et al. \cite{kim2018secure} rely on homomorphic encryption to train an approximated logistic regression function. Zheng et al. \cite{popa2019helen} combine homomorphic encryption and distributed convex optimization, in their system called Helen, in order to collaboratively train linear models. Recently, multiple solutions based on federated learning (relying on differential privacy and edge computing) have been proposed \david{\cite{shokri2015privacy, abadi2016deep, geyer2017differentially, chaudhuri2009privacy, pathak2010multiparty,kim2019secure, jayaraman2018distributed}. These solutions aim at protecting the resulting model from inference attacks~\cite{fredrikson2015model,shokri2017membership}. Some of these works ~\cite{abadi2016deep, chaudhuri2009privacy} assume a trusted party that holds the data, trains the machine-learning model, and performs the noise addition to achieve differential privacy guarantees. Other works \cite{shokri2015privacy, du2018privacy, geyer2017differentially,  boyd2011distributed,huang2019dp} propose solutions for distributed settings in which the parties exchange differentially private model parameters with the help of an untrusted server that trains a collective global model. These approaches are computationally efficient but usually require very high privacy budgets to obtain a useful collective model (due to the noise addition); hence it is unclear what privacy protection they achieve in practice~\cite{jayaraman2019evaluating}. To this end, some works attempt to obtain more useful models in the distributed setting by combining differential privacy with homomorphic encryption~\cite{pathak2010multiparty,kim2019secure} or multi-party computation techniques~\cite{jayaraman2018distributed}. However, most of these solutions are specifically tailored, parameterized and optimized for a given operation, e.g. gradient descent, and would require a redesign if used for different operations. Finally, they assume a weaker threat model with honest-but-curious computing parties and, unlike \sys, they do not enable verification of computation correctness and results robustness.} 

%% file: background.tex
%\vspace{-0.5em}
\section{Background}\label{sec:back}
We introduce \sys's main components and two exemplifying use cases. We describe the cryptographic tools that we use to distribute trust and workload. We present the blockchains \david{that we use to implement our solution} to ensure \sys's correctness and auditability. Finally, we introduce the notion of differential privacy and verifiable shuffle, which are at the core of our solution to ensure individuals' privacy.
\vspace{-0.5em}
\subsection{Use Cases}\label{sub_ch:running_ex}
%In today's world, personalized medicine is considered to be the future of the health industry. However, this new paradigm requires the sharing of data, both medical and genomic, among institutions. Multiple initiatives have been put in motion, in the last coupe of years, to lay down the foundations for personalized medicine and data sharing for example the Patient-Centered Clinical Research Network (PCORNet) [1] in the USA, eTRIKS/- TranSMART [2] in the EU, the Swiss Personalized Health Network (SPHN) [3] in Switzerland, and the Global Alliance for Genomics and Health (GA4GH) [4], which are laying down the foundations for what will definitely be a revolution on the way we treat patients and perform medical research. With that in mind, we illustrate the possible use of \sys in 2 specific settings: (1) Hospital Data Sharing ($HDS$): multiple hospitals want to enable statistical computations and the training of machine learning models across their patients' databases, (2) Personal Data Sharing ($PDS$): the goal is to run studies, e.g., on heart issues, by directly computing on data collected from people wearables. We show that \sys is suitable in both settings and ensures a high level of confidentiality and privacy for the patients.
We illustrate \sys's utility in the medical sector, as it is a paradigmatic example where privacy is paramount and data sharing is needed. Recently, multiple initiatives have emerged to realize the promise of personalized medicine and to address the challenges posed by the increasing digitalization of medical data \cite{GA4GH, selby2012patient, sphn}. In this context, the ability to share highly sensitive medical data while protecting patients' privacy is becoming of primary importance. We illustrate the possible use of \sys in two specific settings %that can be easily extrapolated to other conditions and fields 
that cover most medical data sharing scenarios: (1) Hospital Data Sharing ($HDS$), where multiple hospitals enable statistical computations and the training of machine-learning models across their datasets of patients (e.g., \cite{sphn, healthlnk}), and (2) Personal Data Sharing ($PDS$), where a medical institute runs studies, e.g., on heart issues, by directly computing on data collected from people's wearables (e.g., \cite{appleWatch, p4mi}). %We show that \sys is suitable for both settings and ensures a high level of confidentiality and privacy for the patients.
\vspace{-0.5em}
\subsection{ElGamal Homomorphic Encryption}
\label{EGHE}
\sys requires an additively homomorphic cryptosystem; we choose to rely on the Elliptic Curve ElGamal (ECEG) \david{\cite{elgamal1985public}}, which enables an efficient use of zero-knowledge proofs for correctness \david{\cite{camenisch1997proof}}. However, \sys's functionality is not bound to this choice and can be achieved with other cryptosystems. \david{ECEG relies on the difficulty of computing a discrete logarithm in a finite field; in this case, an Elliptic Curve subgroup of $\mathbb{Z}_p$, with $p$ a big prime.} The encryption of a message $m \in \mathbb{Z}_p$ is E$_\Omega$($m$) = $(rB,\ mB+r\Omega)$, where $r$ is a uniformly-random nonce in $\mathbb{Z}_p$, $B$ is a base point on an elliptic curve $\mathcal{G}$ and $\Omega$ a public key. The table of symbols is presented in Appendix \ref{sym_table}. The additive homomorphic property states that E$_\Omega$($\alpha m_1 + \beta m_2$) = $\alpha$E$_\Omega$($m_1$) + $\beta$E$_\Omega$($m_2$) for any messages $m_1$ and $m_2$ and for any scalars $\alpha$ and $\beta$. In order to decrypt a ciphertext $(rB,\ mB+r\Omega)$, the holder of the corresponding private key $\omega$ ($\Omega=\omega B$) multiplies $rB$ and $\omega$ yielding $\omega (rB) = r\Omega$ and subtracts this point from $mB+r\Omega$. The result $mB$ is then mapped back to $m$, e.g., by using a hashtable. \sys relies on fixed-point representation to encrypt floating values.
\vspace{-0.5em}
\subsection{Zero-Knowledge Proofs}\label{sub_ch:zkp_proofs}
Universally-verifiable zero-knowledge proofs (ZKPs) can be used to ensure computation integrity and to prove that encrypted data are within given ranges. In \sys, we choose to verify computation integrity by using the proofs for general statements about discrete logarithms, introduced by Camenisch and Stadler \cite{camenisch1997proof}. These proofs enable a verifier to check that the prover knows the discrete logarithms $y_1$ and $y_2$ of the public values $Y_1 = y_1B$ and $Y_2 = y_2B$ and that they satisfy a linear equation 
\setlength{\belowdisplayskip}{2pt}
\setlength{\abovedisplayskip}{2pt}
\begin{equation}\label{eq:proof_equation}
A_1y_1+A_2y_2 = A,
\end{equation} 
where $A$, $A_1$, $A_2$ are public points on $\mathcal{G}$. This is done without revealing any information about $y_1$ or $y_2$. \\
The input-range validation is done by relying on the proofs proposed by Camenisch and Chaabouni \cite{camenisch2008efficient}, with which we can prove that a secret message $m$ lies in a given range $[0,u^l)$ with $u$ and $l$ integers, without disclosing $m$. \david{The prover writes the base-u decomposition of its secret value $m$ and commits to the u-ary digits by using the verifier signatures on these digits. The $l$ created commitments prove to the verifier that $m \in [0,u^l)$.} We present this proof, adapted to our framework, in Algorithm \ref{alg:the_alg}. Finally, both proofs can be made non-interactive through the Fiat-Shamir heuristic \cite{fiat1986prove}.
\vspace{-0.5em}
\subsection{Interactive Protocols}\label{sub_ch:collective_authority} 
Interactive protocols can be used to distribute the computations and the trust among multiple computing nodes $CNs$. In \sys, each  $CN_i$ possesses a private-public key pair ($k_i$, $K_i$) where $k_i$ is a uniformly-random scalar in $\mathbb{Z}_p$  and $K_i = k_i B$ is a point on $\mathcal{G}$. The $CNs$' collective public key is $K = \textstyle \sum_{i=1}^{\#CN}K_i$. The corresponding secret key $k = \textstyle \sum_{i=1}^{\scriptstyle\#CN}k_i$ is never reconstructed such that a message encrypted by using $K$ can be decrypted only with the participation of all $CNs$. An attacker would have to compromise all $CNs$ in order to decrypt a message. \david{As shown in Section \ref{sec:roadmap}, to produce the intended results, \sys protocols require the participation of all the $CNs$.}
%\vspace{-0.8em}
\subsection{Blockchains}
A \textit{blockchain} is usually a public, append-only ledger that is distributively maintained by a set of nodes and serves as an immutable ledger  \cite{kokoris2016managing, nakamoto2008bitcoin}. \david{Its main applications are in cryptocurrencies \cite{nakamoto2008bitcoin, wood2014ethereum} but is also used in other domains, e.g., health care \cite{kuo2017blockchain}. Data are bundled into blocks that are validated through the consensus \cite{castro1999practical, yin2018hotstuff} of the maintaining nodes. Each block contains a pointer (i.e., a cryptographic hash) to the previous valid block, a timestamp, a nonce, and application-specific data. The chain of these blocks forms the blockchain.}
\vspace{-0.5em}
\subsection{Differential Privacy}\label{sub_ch:diffP}
Differential privacy is an approach for privacy-preserving reporting results on statistical datasets, introduced by Dwork~\cite{df}. This approach guarantees that a given randomized statistic, $\mathcal{M}(DS)=R$, computed on a dataset $DS$, behaves similarly when computed on a neighbor dataset $DS'$ that differs from $DS$ in exactly one element. More formally, ($\epsilon$, $\delta$)-differential privacy \david{\cite{dwork2006calibrating}} is defined by $\Pr \left[\mathcal{M}(DS) =R\right] \leq \exp(\epsilon) \cdot \Pr \left[\mathcal{M}(DS')=R\right] + \delta$, where $\epsilon$ and $\delta$ are privacy parameters: the closer to 0 they are, the higher the privacy level is. \david{($\epsilon$, $\delta$)-differential privacy is often achieved by adding noise to the output of a function $f(DS)$. This noise can be drawn from the Laplace distribution with mean 0 and scale $\frac{\Delta f}{\epsilon}$, where $\Delta f$, the sensitivity of the original real valued function $f$, is defined by $\Delta f =\max_{D,D'}{||f(DS)-f(DS')||_1}$. Other mechanisms, e.g., relying on a Gaussian distribution, have also been proposed \cite{ghosh2012universally, dwork2014algorithmic}.}

\vspace{-0.5em}
\subsection{Verifiable Shuffles}\label{sub_ch:verifiable_shuffle}
To randomly select a value from a public list of noise values and to ensure differential privacy, we \david{rely on a verifiable shuffle \cite{nefffirstverifiable, neffverifiable, bayer2012efficient, groth2003verifiable}. We implemented and use the verifiable shuffle} of ElGamal pairs, described by Neff \cite{neffverifiable}. This protocol takes as input a list of $\chi$ ElGamal pairs $(C_{1,i},\ C_{2,i})$ and outputs $(\bar{C}_{1,i},\ \bar{C}_{2,i})$ pairs such that for all $1 \leq i \leq \chi$, $(\bar{C}_{1,i}, \bar{C}_{2,i}) =(C_{1,\upsilon(i)} + r''_{\upsilon(i)}B, C_{2,\upsilon(i)}+ r''_{\upsilon(i)}\Omega)$, where $r''_{\upsilon(i)}$ is a re-randomization factor, $\upsilon$ is a permutation and $\Omega$ is a public key. \david{The permutation $\upsilon$ is used to change the order of the ElGamal pairs and $r''_{\upsilon(i)}$ is used to modify the value of the ciphertext encrypting a message $m$ such that its decryption still outputs $m$. As a result, an adversary not knowing the decryption key $\upsilon$ and the $r''_{\upsilon(i)}$ is unable to link back any ciphertext $(\bar{C}_{1,i}, \bar{C}_{2,i})$ with a ciphertext $(C_{1,i},\ C_{2,i})$.} Neff provides a method for proving that such a shuffle is done correctly, i.e., that there exists a permutation $\upsilon$ and re-randomization factors $r''_{i,j}$ such that $output$ = $\scriptstyle{\textit{SHUFFLE}}$$_{\upsilon,r''_{i,j}}$$(input)$, without revealing anything about $\upsilon$ or $r''_{i,j}$. This is achieved by using honest-verifier zero-knowledge proofs, introduced by Neff \cite{nefffirstverifiable, neffverifiable}.

%% file: model.tex
\section{System Overview}\label{sec:arch}
In this section, we describe the system and threat models, before presenting \sys's functionality and security requirements.
\vspace{-0.5em}
\subsection{System Model}\label{sec:sysModel}
\begin{figure}[h!]
	\vspace{-1.0em}
	\centering
	\includegraphics[width=0.6\columnwidth]
	{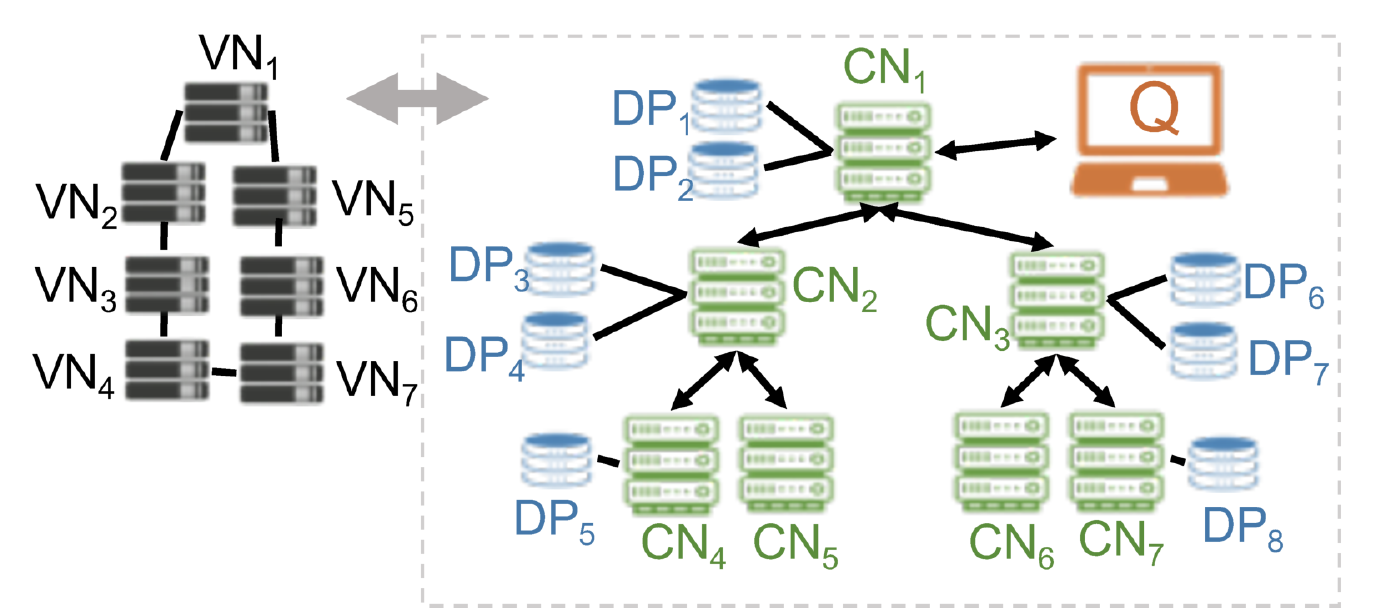}
	\vspace{-0.5em}
	\caption{A querier $Q$, Data Providers $DP_i$, Computing Nodes $CN_i$ and Verifying Nodes $VN_i$.}
	\label{fig:sys_mod}
	\vspace{-0.5em}
\end{figure}

The system model is represented in Figure \ref{fig:sys_mod}. \david{For simplicity, we describe here the logical roles in \sys, and in Section \ref{sec:discussion} we discuss the fact that a physical node can simultaneously play multiple roles.}
A querier $Q$ can execute a statistical query and the training and evaluation of a machine-learning model on distributed datasets held by $DPs$.
\david{The $CNs$ collectively handle the computations in the system; i.e., from $Q$'s perspective, they emulate a central server and provide answers to her queries. The verifying nodes' ($VNs$) role is to provide auditability; they collectively verify the query execution and immutably store the corresponding proofs. They enable an auditor, e.g., $Q$ or an external entity, to easily verify (audit) the correctness of the query execution.}

\david{In \sys's typical workflow, }the query is defined by the querier $Q$ and is then broadcast to the $CNs$ and $DPs$. The $DPs$ answer with their encrypted responses that are then collectively aggregated and processed by the $CNs$, before the result is sent to $Q$. \david{We assume that the used data formats are sufficiently homogeneous among different $DPs$ and that the $DPs$ are able to interpret the queries, e.g., there is a common ontology of attributes and the query-language is agreed-on during system setup}.

An exemplifying instantiation of this system model in the $HDS$ scenario (Section \ref{sub_ch:running_ex}) would feature the $CNs$ as universities that want to enable researchers ($Q$) to compute on data held by multiple hospitals ($DPs$). $VNs$ can be independent or governmental institutions ensuring that data protection regulations are respected.

We assume that the system's topology and public information, e.g., public keys, are known by all entities. Authentication and authorization are out of scope of this paper and we briefly discuss them in Section \ref{sec:discussion}.
%We notice that a node in the system can play multiple roles, as nothing prevents a $DP$ or $CN$ to also be a $VN$ and thus participate in the query verification. In our running examples, in $HDS$, a hospital can be a $DP$, a $CN$ and even a $VN$, while in $PDS$, an individual is a single $DP$. In both cases, the $VNs$ can be independent or governmental institutions.

%\vspace{-0.8em}
\subsection{Threat Model}\label{sec:threatfunctionalityModel}

\noindent We assume a strong threat model:
\begin{itemize}[leftmargin=*]
	\item \emph{Queriers}. They are considered malicious as they can try to infer information about the $DPs$ from the queries end results or by colluding with other entities in the system.
	\item \emph{Computing Nodes}. We consider an Anytrust model \cite{wolinsky2012scalable}, which means that all \sys's security and privacy guarantees (Section \ref{sec:secuRequirements}) are ensured, as long as at least one of the $CNs$ is honest-but-curious (or plain honest).
	\item \emph{Data Providers}. The $DPs$ are considered malicious as they can try to produce an incorrect answer to a query in order to bias the final results. They can also collude with other nodes to infer information about other $DPs$ or about a query end-results.
	\item \emph{Verifying Nodes.} We assume that a threshold number of the $VNs$ is honest. This threshold, e.g., $f_h = 2f+1$ out of $f_t = 3f+1$, where $f_t$ is the number of $VNs$, is defined depending on the consensus algorithm \david{\cite{castro1999practical, yin2018hotstuff}} that is used to ensure \david{a correct and immutable storage of the proofs' verification results.} 
	%For example, with the PBFT algorithm \david{\cite{castro1999practical, yin2018hotstuff}}, the threshold is $f_h = 2f+1$ out of $f_t = 3f+1$, where $f_t$ is the number of $VNs$. 
\end{itemize}
%\vspace{-0.3em}
\subsection{Functional Requirements}\label{sec:func}
%\vspace{-0.3em}
\sys enables the computation on distributed datasets of any operation in the family of \textit{encodable operations}. An \textit{encodable operation} can be separated in two parts: the $DPs$' local computations and the collective aggregation. In the collective part, the computations are executed on encrypted data and are thus limited by the homomorphism in the used cryptographic scheme, e.g., additions and/or multiplications. $DPs$' computations are executed locally and are therefore not limited.
%\david{division btw. local (non restricted encoding) and collective part -> no leakage -> HE -> depends on cryptoscheme, e.g. additive, linear or affine, can be polynomial if SHE}
\vspace{-0.2em}
\david{
\begin{definition}
	An \emph{encodable operation} $f$ computed among $N$ $DPs$ is defined by:
	\[
	%f(\{\bar{r_i}\}_{i=1}^N) \equiv \pi(\{\rho(\bar{r_i})\}_{i=1}^N)
	f(\bar{r}) \equiv \pi(\{\rho(\bar{r_i})\}_{i=1}^N),
	%f(\bar{r}) \equiv \pi(\sum_{i=1}^{N}\{\rho(\bar{r_i})\}),
	\]
	in which the \emph{encoding} $\rho$ is defined by
	\[
	\rho(\bar{r_i})\equiv (\mathbf{V_i},c_i),
	\]
	where $\mathbf{V_i}=[v_{i,1},...,v_{i,d}]$ is a vector of $d$ values computed on a set of $c_i=|\bar{r_i}|$ records, where $|.|$ stands for cardinality. $\bar{r}$ is the set of all distributed datasets' records, $\bar{r_i}$ is the set of records that belong to $DP_i$, and $\pi$ is a polynomial combination of the outputs of the encodings $\rho$. The encodings are defined as locally computed functions on the subsets ($\bar{r_i}$) of each $DP_i$. It is also possible to express an encodable operation as a recursive function:
	\[
	f_k(\bar{r}) \equiv \pi(\{\rho(\bar{r_i}, f_{k-1}(\bar{r}))\}_{i=1}^N).
	\]
	%In other words, for a specific operation $f$, each $DP_i$ creates an encoding $\rho$ of its local result computed on its set of records $\bar{r_i}$. These encodings are aggregated among all $DPs$ by the $CNs$ and the querier performs $\pi$ on the aggregated result.
	%
	%
	%where $\bar{r}$ is the set of all distributed datasets' records, $\bar{r_i}$ is a set of records belonging to a $DP_i$ and $\pi$ is a linear combination of the outputs of the encodings $\rho$. In other words, for a specific operation $f$, each $DP_i$ creates an encoding $\rho$ of its local result computed on its set of records $\bar{r_i}$. An \emph{encoding} $\rho$ is defined by
%	\[
%	\rho(\bar{r_i})\equiv (\mathbf{V_i},c_i),
%	\]
	%where $\mathbf{V_i}=[v_{i,1},...,v_{i,d}]$ is a vector of $d$ values computed on a set of $c_i=|\bar{r_i}|$ records, where $|.|$ stands for cardinality.
	%The encodings $ \rho(\bar{r_i})$ produced by the $DPs$ are linearly combined, as defined by $\pi$, to compute $f$.
	%
	%
	%\david{An operation $\psi$, e.g. variance, is executed on a set of $DPs$ in which each $DP_i$ produces a vector $\mathbf{V_i}$ by computing on $c_i$ records $r$ in its local database. An \enco is defined by  \{$\psi$, $\mathbf{V}=[v_1, ..., v_d]$, $c$\} where $\mathbf{V}$ and $c$ are the sums of all $\mathbf{V_i}$ and $c_i$, respectively. $\psi$ defines how the elements in $\mathbf{V}$ and $c$ are combined to compute the operation's result.}
	\label{def:encoding}
\end{definition}
%\vspace{-0.2em}
}
\david{In \sys, for any specific operation $f$, each $DP_i$ creates an encoding $\rho$ computed on its set of records $\bar{r_i}$. Then, $\pi$ is executed in two parts: the $CNs$ first aggregate all $DPs$' encodings outputs ($\sum_{i=1}^{N}\{\rho(\bar{r_i})\}$) and, if needed, the querier post-processes $\pi$ on the aggregated result (e.g, if $\pi$ involves information-preserving operations not executable by the CNs under homomorphic encryption).}

\david{We give here an instantiation of Definition \ref{def:encoding} that enables the computation of the \texttt{average}, and in Section \ref{sec:encoding} we show how an \enco can be instantiated to enable the computation of: \texttt{sum}, \texttt{count}, \texttt{frequency count}, \texttt{average}, \texttt{variance}, \texttt{standard deviation},  \texttt{cosine similarity}, \texttt{min/max}, \texttt{AND/OR} and \texttt{set intersection/union}, and the training and evaluation of \texttt{linear} and \texttt{logistic regression} models.}

\david{For example, if $Q$ wants to compute the \texttt{average} ($f$) heart rate over multiple patients across hospitals ($HDS$ (Section \ref{sub_ch:running_ex})), each hospital ($DP_i$) answers with the encoding of its (encrypted) local sum of each patient's heart rate ($h$): $\rho(\bar{r_i}) \equiv ([\sum_{j=1}^{c_i}h_{i,j}],c_i)$. These encodings are then (homomorphically) added across all hospitals, and $Q$ can (decrypt and) compute the global average by using $\pi = \sum_{i=1}^Nv_{i,1}/\sum_{i=1}^Nc_i$. We remark here that whereas $\rho$ and $\pi$ are application dependent, the workflow is common to all the possible operations.}

%In Section \ref{sec:encoding}, we show how an \enco can be instantiated to enable the computation of: \texttt{sum}, \texttt{count}, \texttt{frequency count}, \texttt{average}, \texttt{variance}, \texttt{standard deviation},  \texttt{cosine similarity}, \texttt{min/max}, \texttt{AND/OR} and \texttt{set intersection/union}, and the training and evaluation of \texttt{linear} and \texttt{logistic regression} models.
%\vspace{0.5em}
%
Finally, in \sys, an auditor can efficiently audit a query execution. Moreover, the proofs required for auditability are produced such that their creation does not affect the query runtime.
%Operation $f$ is 
%
%\sys works independently of the query language and we rely on SQL to describe its capabilities. \sys is able to answer any %SQL query of the form:
%
%\begin{alltt}
%	SELECT OPERATION \(...\) FROM \(DP\sb{1}, ..., DP\sb{n}\) WHERE/LIKE/ANY\(...\) \(...\)
%\end{alltt}
%where $operation$ belongs to the set of \textit{encodable operations}:
\vspace{-0.2em}
\subsection{Security Requirements}\label{sec:secuRequirements}
%\vspace{-0.3em}
\sys must ensure:
\begin{itemize}[leftmargin=*]
	\setlength\itemsep{0em}
	\item \emph{Data confidentiality}. The data input by the $DPs$ have to remain confidential at any time. Only $Q$ is able to see the query answer.
	\item \emph{$DPs$' privacy}. No entity is able to infer information about one single $DP$ or about any individual storing his data in a $DP$'s database.
	\item \emph{Query Execution Correctness.} We consider the query execution to be correct when both \textit{results \robust} and \textit{computation correctness} requirements are met:
	\begin{itemize}[leftmargin=*]
		\setlength\itemsep{0em}
	\item \emph{Results \robust}. The query results are protected against strong outliers, either maliciously or erroneously input by the $DPs$. %\david{REMOVE: I would remove all the rest (here we just list the requirements no need to analyse) Malicious $DPs$ cannot send specifically crafted values, e.g., extremely high values, to bias the result of a query above a certain limit. $DPs$ can still input incorrect values, but their influence on the final result is limited. We give an intuition on how malicious $DPs$ are prevented from highly distorting results in Section \ref{prototype_evaluation}. We remark that this requirement also ensures that an outlier, e.g., a $DP$ inputting an erroneous value, is either part of the range and has a limited impact on the statistic, or creates a wrong range proof.}
	\item \emph{Computation correctness.} Any computation undertaken by the $CNs$ is correctly executed.
	\end{itemize}
\end{itemize}

%% file: design.tex
\section{Drynx Design}\label{sec:roadmap}
%To the best of our knowledge, most of the related works assume $DPs$ to be honest-but-curious and to send correct data. The only exception is Prio \cite{corrigan2017prio}: it provides result robustness against $DPs$ sending incorrect values. However, in Prio, this is only possible when all the $CNs$ are honest. Moreover, Prio does not protect against $DPs$ colluding among themselves or with $CNs$. Finally, no previous work proposes a practical solution for computation correctness and auditability. 
To overcome the limitations in existing works and meet the requirements presented in the previous section, we propose a novel system model in which we enable query auditability by introducing $VNs$. Additionally, \sys provides multiple functionalities in a stronger threat model by relying on $DPs$ that encode locally computed results proven to be within a certain range. It limits the trust in $DPs$ by controlling that their results are in these pre-defined ranges. We propose a system that remains generic and practical while operating in a threat model stronger than existing works. We discuss now the design of this system.
%\joao{In our previous work, UnLynx, we designed a system enabling the computation of sums across multiple $DPs$. We considered the $CNs$ to follow an Anytrust model, but assumed the $DPs$ to be honest and to send correct data. Moreover, the $CNs$ had to prove the correctness of their computations, but we didn't tackle the problem of how to store and verify these information. 
%\david{We remark that unlike related works (the closest being UnLynx), \sys has to provide query execution correctness in a threat model in which no entity is individually trusted. Additionally, \sys provides multiple functionalities, whereas UnLynx only enables sums. This is achieved in our strong threat model by relying on $DPs$ computing on their local data and encoding their results. \sys limits the trust in $DPs$ by controlling that their results are in specific ranges. In order to enable auditability and an efficient verification of \sys's behavior we introduce $VNs$in our novel system model. We therefore propose a system that remains generic and practical while operating in a threat model stronger than existing works.}

In \textit{\sys's Security Design} (Section \ref{sec:security}), we show how we build \sys to meet all its security requirements: %We introduce a simple construction of \sys that provides data confidentiality and describe how to build upon to meet the remaining requirements.
\begin{itemize}[leftmargin=*]
	\item In Section \ref{subsec:dataconf}, we introduce a simple query-execution pipeline enabling \sys's functionalities and protecting data confidentiality.
	\item In Section \ref{subsec:dpPrivacy}, we build upon the previously introduced query-execution pipeline and explain how to ensure $DP$s' privacy by introducing the new concept of a \textit{neutral encoding}. This enables a $DP$ to privately choose whether to answer a query. We also explain how \sys handles bit-wise operations and maintains $DPs$' privacy. Finally, we introduce distributed differential privacy that is used to ensure that no entity infers information about a single $DP$ or individual from the query end results.
	\item In Section \ref{subsec:queryCorrect}, we show how we provide auditability in an efficient way by relying on a set of $VNs$. We describe how \sys ensures results robustness by leveraging on range proofs and how all \sys's computations can be verified by relying on proofs of correctness.
\end{itemize}
In \textit{\sys's Optimized Design} (Section \ref{sec:performance}), we discuss how to optimize \sys's performance:
\begin{itemize}[leftmargin=*]
	\item In Section \ref{subsec:parallelization}, we present \sys's full query-execution pipeline. We show how multiple parts of the query execution and verification can be run concurrently thus optimize \sys's runtime.
	\item In Section \ref{subsec:probaVerif}, we introduce a tradeoff between security and performance by enabling a probabilistic verification of the query execution.
\end{itemize}
%\vspace{-0.4em}
\subsection{Drynx Security Design}\label{sec:security}
We present \sys core security architecture.% by describing how data confidentiality, $DPs$ privacy, and query-execution correctness are ensured.

\vspace{0.4em}
\subsubsection{\bf Data Confidentiality}\label{subsec:dataconf}
\begin{figure}[h!]
	%\vspace{-0.5em}
	\centering
	\includegraphics[width=0.4\columnwidth]
	{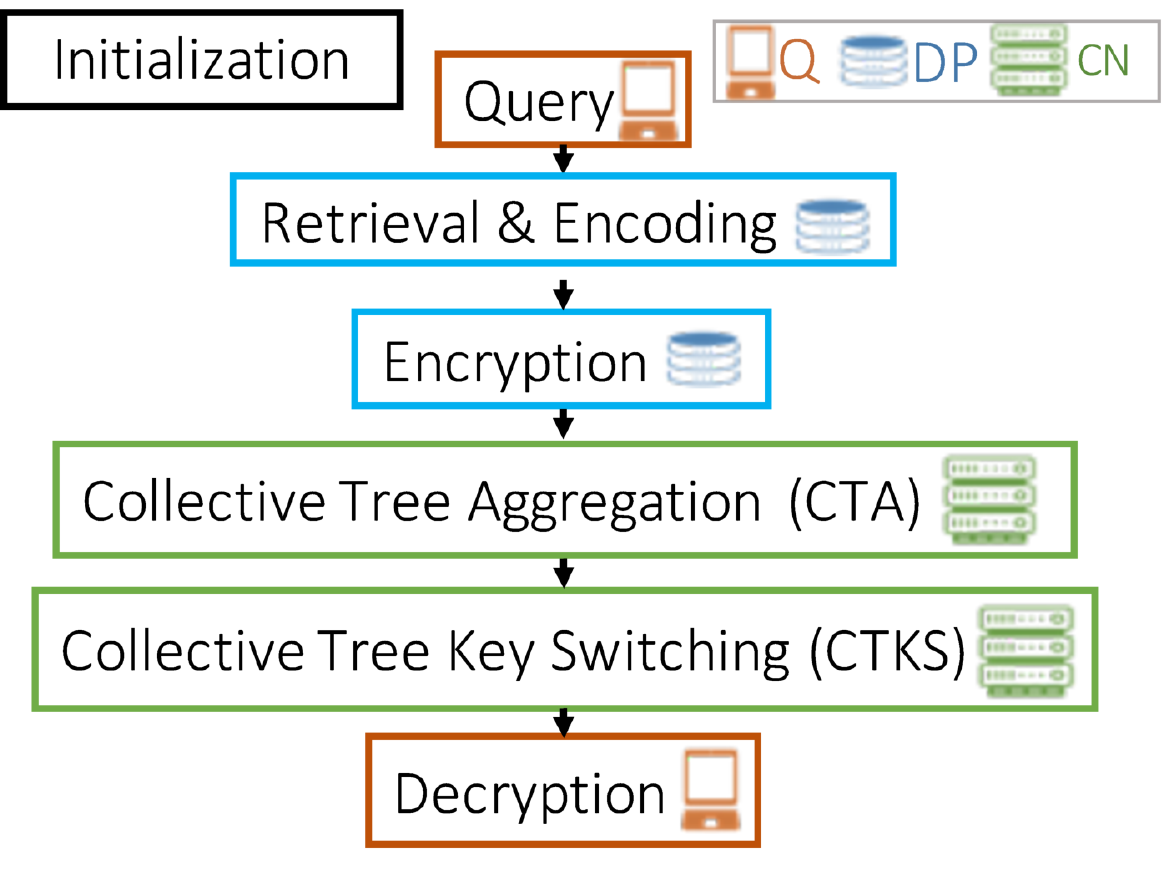}
	\vspace{-0.6em}
	\caption{Confidential System Query Execution.}
	\label{fig:strynx}
	%\vspace{-0.5em}
\end{figure}

First, we introduce a confidential distributed data-sharing system (Figure \ref{fig:strynx}) that can run the same operations as \sys, but only meets one of the security requirements: \textit{data confidentiality}. %$CNs$ computations and the $DPs$' answers are not verified, and the $DPs$' privacy is not protected.

 We describe the query execution protocol, and sketch the proof of confidentiality for this system. Afterwards, we describe how to enhance this construction to meet \sys's other security requirements without breaking data confidentiality.
\begin{enumerate}[leftmargin=*]
	%\vspace{-0.5em}
	%\setlength\itemsep{0em}
	\item  \emph{Initialization.} Each $CN_i, DP_i$ and $Q_i$ generates its own private-public key-pair $(k_i, K_i)$. The $CNs$' public keys are then summed up in order to create $K$, the $CNs$' public collective key that is used to encrypt all the processed data. %Whenever a new $CN$ joins or leaves the system, this key is updated accordingly.
	\vspace{0.2em}
	\item \emph{Query.} $Q$ formulates the query that is broadcast in clear through the $CNs$ to the $DPs$. \david{Although the querier could directly communicate with the $DPs$, our choice simplifies the communication scheme and the synchronisation inside the system, as the $CNs$ have to know the query and receive the $DPs$ inputs to perform the computations in the remaining steps.} 
	%The querier sends the query through the $CNs$ as those ones have to know the query to perform the computations in the remaining steps. Moreover, this simplifies the communication scheme and the synchronisation inside the system. $Q$ only communicates with one $CN$ and in the following steps, each $CN$ has to handle the communications and to wait for the $DPs$' answers before proceeding with the remaining parts of the query execution.}
	The query defines the operation, the attributes on which the operation is computed, the participating $DPs$ and (optionally) the filtering conditions. \sys works independently of the query language. We illustrate its use with a SQL-like query to compute the \texttt{average} heart rate among patients for which data are held by $n$ $DPs$:
	\begin{alltt}
		SELECT \texttt{average} \(heart\_rate\) ON \(DP\sb{1}, ..., DP\sb{n}\)
		WHERE \(patient\_state = 'hypertensive'\)
	\end{alltt}
	\vspace{0.2em}
	\item \emph{Retrieval \& Encoding}. The $DPs$ compute their local answer by following $\rho$ which is defined in the operation \enco (Definition \ref{def:encoding}). For this purpose, they first locally retrieve the corresponding data.%\david{\sys is agnostic to the $DP$'s local-database engine. For example, if a $DP$ relies on SQL to retrieve the data from its table $R$, it can run:
	%\begin{alltt}
	%	SELECT (\(R.heart\_rate\))  FROM \(R\)
	%	WHERE \(R.patient\_state = 'hypertensive'\)
	%\end{alltt}}
	\vspace{0.2em}
	\item \emph{Encryption}. The $DPs$ encrypt their encoded answer under $K$ and send the corresponding ciphertexts back to the $CNs$. 
	\vspace{0.2em}
	\item \emph{Collective Tree Aggregation ($CTA$)}. The $CNs$ collectively aggregate all $DPs$' responses by executing a $CTA$ protocol relying on the \emph{Collective Aggregation} protocol defined in UnLynx \cite{froelicher2017unlynx}. The $CNs$ are organized into a tree structure such that each $CN$ waits to receive the aggregation results from its children and sums them up before passing the result on to its own parent.
	\vspace{0.2em}
	\item \emph{Collective Tree Key Switching ($CTKS$)}. The $CNs$ collectively convert the aggregated result, encrypted under $K$, to the same result encrypted under $Q$'s public key $K'$, without ever decrypting. This protocol (Protocol \ref{CTKS}) is a new construction of the \emph{Key Switching} proposed in UnLynx \cite{froelicher2017unlynx}. \david{Conceptually, each $CN$ partially decrypts $m$ (i.e., the term $- (C_1)k_i$ in the computation in step 2) and re-encrypts it with $Q$'s public key $K'$ (i.e., the term $+ \alpha_{i}K'$ in step 2).}
	\vspace{0.5em}
	\begin{protocol}{Collective Tree Key Switching (CTKS)}
		\addtocounter{protocol}{-1}
		\refstepcounter{protocol}
		\label{CTKS}
		\vspace{-0.5em}
		\setlength\itemsep{0em}
		\small
		\textit{Input.} E$_K(m$) $= (C_{1},\ C_{2}) = (rB, mB+rK),\ K'$\\
		\small
		\textit{Output.} E$_{K'}(m$) $=\scriptsize({C'}_{1},{C'}_{2}) =  (r'B, mB+r'K')$\\
		\vspace{-0.5em}
		\small
		\textit{Protocol.}
		\begin{enumerate}[1.]
			\setlength\itemsep{0em}
			\small
			\item The root $CN_1$ sends $C_{1}$ down the tree to all $CNs$.
			\item Each $CN_i$ generates a secret uniformly-random nonce $\alpha_i$ and computes  $w_{i,1} = \alpha_{i}B$ and $w_{i,2} = - (C_1)k_i + \alpha_{i}K'$
			\item The $CNs$ collectively aggregate (i.e., using $CTA$) all the $w_{i,1}$ and  $w_{i,2}$.
			\item $CN_1$ finally computes $\textstyle({C'}_{1},{C'}_{2}) = (\textstyle\sum  w_{i,1},C_{2} + \textstyle\sum  w_{i,2}) =  (r'B, mB+r'K')$ where $r' = \textstyle\sum \alpha_i$.
			\vspace{-1.0em}
		\end{enumerate}
	\end{protocol}
	\vspace{0.5em}
	We improve the efficiency of $CTKS$ by changing the way the ciphertexts are transformed and by organizing the $CNs$ in a tree structure, \david{thus reducing its execution time. In this structure, multiple $CNs$ can perform their local operations (3 scalar multiplications and 1 addition) in parallel, and the $CTA$ requires $\#CN-1$ aggregations and communications between the nodes. We show the computational complexity of all \sys protocols in Table \ref{tab_band}.}
	
	\item \emph{Decryption.} $Q$ decrypts and decodes the query results.
\end{enumerate}
\vspace{0.5em} 
\emph{Security Arguments.} We show that, as long as one $CN$ is honest, an adversary who controls the remaining $CNs$, $DPs$ and $Q$ cannot break data confidentiality. Without loss of generality, we assume that at least one $DP$ is honest, as only in this case there is data to protect from the adversary. We sketch the proof by relying on the real/ideal simulation paradigm \cite{lindell2017simulate} and show that an adversary cannot distinguish a ``real'' world experiment, in which the adversary is given ``real'' data (sent by honest $DPs$), and an ``ideal'' world experiment, in which the adversary is given data (e.g., random) generated by a simulator.  It can be shown that the $DPs$ send encrypted data that are never decrypted before being aggregated and re-encrypted ($CTKS$) under $Q$'s public key. Therefore, due to the cryptosystem's semantic security, the adversary cannot distinguish between a simulation and a real experiment. It can be seen that data confidentiality is thus ensured during end-to-end query execution:
%\david{\textbf{Confidentiality}. JUAN CAN YOU REWRITE THIS PARAGRAPH? We rely on the Real/Ideal Simulation Paradigm \cite{lindell2017simulate} to sketch the proof for confidentiality. In the ideal world simulation, a trusted party plays the $CNs$' role. It receives $DPs$' answers and computes the query's result such that $Q$ is the only entity to see the result's \enco. We hereby show that our simplified confidential system achieves this functionality in the real world scenario. We show in the following that due to the cryptoscheme's semantic security and since $CNs$ only see encrypted data, this can be simulated with random inputs. The corresponding simulator is therefore straightforward to build.}

In \emph{Retrieval \& Encoding}, the $DPs$ operate only on their local data and no external data is seen by any malicious party. In \emph{Encryption}, the $DPs$ encrypt their responses with $K$ and these responses are aggregated, still under encryption, in $CTA$. The (summed) ciphertexts cannot be decrypted unless all $CNs$ collude, which is not possible as they follow an Anytrust model. Finally, in $CTKS$ (Protocol \ref{CTKS}), a ciphertext is switched from $K$ to $Q$'s public key such that $Q$ can decrypt: 
\begin{itemize}[leftmargin=*]
    \setlength\itemsep{0em}
    \item in \textit{$CTKS$ Steps: 1-3. } The ciphertext is encrypted under $K$ and thus cannot be decrypted without the collusion of all $CNs$.
    \item in \textit{$CTKS$ Step: 4. } The ciphertext is always $\textstyle(\tilde{C}_{1},\tilde{C}_{2}) = (\tilde{r}B, mB+\tilde{r}K')$ where $\tilde{r} = \textstyle\sum_{i=0}^{t} \alpha_i$ and $0 \leq t \leq \#CN$ and can only be decrypted if the $t$ $CNs$ collude with $Q$, who is the intended recipient of the message.
\end{itemize}
\vspace{0.5em}
\subsubsection{\bf DPs' Privacy}\label{subsec:dpPrivacy}
\sys protects $DP$s' and individuals' privacy by ensuring that (a) each $DP$ can privately decide whether to answer a query, (b) only the result of the operation, as defined by the operation \enco, is disclosed to $Q$, and (c) no entity can infer information about a single $DP$ or individual.
\vspace{0.3em}
\noindent \paragraph{\bf Neutral Response}
If a $DP$ determines that a query can jeopardize its privacy, it can choose to not respond, or answer with a neutral response, thus hiding its refusal to participate in the query without distorting the query results. For this purpose we define \nenco:
\begin{definition}
	%\vspace{-0.5em}
	A $DP_i$ sends a \emph{neutral response} by defining its response \enco (Definition \ref{def:encoding}) by $\rho(\bar{r_i})\equiv (\mathbf{O},0)$, where $\mathbf{O}$ is the neutral vector such that $\mathbf{W} + \mathbf{O} = \mathbf{W}$ with $\mathbf{W}$ being any \enco vector; $c_i=0$ as $DP_i$ computes on 0 records.
	%\vspace{-0.5em}
\end{definition}
In Section \ref{sec:encoding}, we describe how a \nenco can be generated for each listed \enco.

\emph{Security Arguments.} A $DP$ not answering a query would suggest (leak) to other entities that this query is too sensitive for it. $DPs$' responses are always encrypted and, due to the indistinguishability property of the underlying cryptosystem, a \nenco is indistinguishable from a non-neutral one, thus effectively hiding the $DP$'s refusal. %\david{REMOVE ? Moreover, a neutral response does not distort the query result and $Q$ knows the number of samples considered in the computation ($c$) as it is part of the \enco.}
\vspace{0.3em}
\paragraph{\bf Privacy-Preserving Bit-wise Operations}\label{par:bitwise}
In \sys, $DPs$' responses are summed through the available additive homomorphism; if these responses are binary, the result of the sum can leak to $Q$ more than the operation result. For example, when an \texttt{OR} operation is executed over a set of $DPs$, $Q$ should only know if the answer is $true$ (1) or $false$ (0). Nevertheless, if the $DPs$' responses are naively summed, $Q$ gets the number of $DPs$ that answered `1' and `0'. To overcome this issue, we propose the \textit{Collective Tree Obfuscation ($CTO$)} protocol, detailed in Protocol \ref{CTO}. For bit-wise operations, $CTO$ is run between steps $CTA$ and $CTKS$ of the query execution. In $CTO$, the $CNs$ collectively obfuscate a ciphertext by multiplying it with a random secret. 

$CTO$ enables privacy-preserving bit-wise operations in \sys as a `1' is obfuscated to a random value whereas `0' is preserved. To know the result of the operation, $Q$ only checks if the final value is `0' or not.%, i.e., he does not have to  the actual value.

%\vspace{-0.2em}
\begin{protocol}{Collective Tree Obfuscation (CTO)}
	\addtocounter{protocol}{-1}
	\refstepcounter{protocol}\label{CTO}
	\vspace{-0.5em}
	\small
	\textit{Input.} E$_K(m$) $= (C_{1},\ C_{2}) = (rB, mB+rK)$\\
	\small
	\textit{Output.} E$_K(sm$) = $(srB, smB+srK)$\\
	\vspace{-0.5em}
	\small
	\vspace{-0.5em}
	\textit{Protocol.}
	\begin{enumerate}[1.]
		\small
		\setlength\itemsep{0em}
		\item Root $CN_1$ sends $(C_{1},\ C_{2})$ down the tree to all $CNs$.
		\item Each $CN_i$ generates a secret uniformly random nonce $s_i$ and computes  $({\hat{C}}_{i,1}, {\hat{C}}_{i,2})= s_i \cdot (C_{1},\ C_{2})$
		\item The $CNs$ collectively aggregate (i.e., using $CTA$) all the $({\hat{C}}_{i,1}, {\hat{C}}_{i,2})$.
		\item $CN_1$ obtains E$_K(sm$) =  $s \cdot (C_{1},\ C_{2})$ where $s = \scriptsize\sum s_i$.
		\vspace{-1.2em}
	\end{enumerate}
	%\vspace{-1.0em}
\end{protocol}

\emph{Security Arguments.}  Protocol \ref{CTO} does not hinder the confidentiality of $m$ and indeed obliviously and statistically obfuscates $m$. \david{The confidentiality relies on the cryptosystem's semantic security, as $m$ remains encrypted during the whole protocol execution. A multiplicative blinding of $m$ in $\mathbb{Z}_p$ is defined by $s \cdot m$, where $s$ is a secret scalar value in $\mathbb{Z}_p$. The output of the $CTO$ protocol is the encryption of $(\scriptsize\sum s_i) \cdot m$. We can rewrite $(\scriptsize\sum s_i) \cdot m$ by separating the contributions of the honest $CNs$ $h$ (at least one $CN$ due to our Anytrust model assumption) and malicious $CNs$ $e$:  $(\scriptsize\sum_{i \in h} s_i + \scriptsize\sum_{i \in e} s_i) \cdot m = (\scriptsize\sum_{i \in h} s_i) \cdot m + (\scriptsize\sum_{i \in e} s_i) \cdot m$. Even if an adversary knows $(\scriptsize\sum_{i \in e} s_i) \cdot m$, the other term $(\scriptsize\sum_{i \in h} s_i) \cdot m$ ensures a multiplicative blinding of $m$ in $\mathbb{Z}_p$.} 
%First, E$_K(m$) is simply multiplied by a uniformly-random scalar, and the confidentiality comes directly from the cryptosystem semantic security. Second, the $CNs$ follow an Anytrust model, i.e., at least one $CN$ does multiply E$_K(m$) with a secret scalar chosen in $\mathbb{Z}_p$, thus ensures a multiplicative blinding of $m$.

\vspace{0.5em} 
\paragraph{\bf Distributed Differential Privacy}
\sys relies on the Collective Differential Privacy ($CDP$) protocol, introduced in Unlynx \cite{froelicher2017unlynx}, to ensure differential privacy, and prevent information inference about some $DPs$ and/or individuals from the query results. For completeness, we briefly present the $CDP$ (Protocol \ref{CDP}) and refer to \cite{froelicher2017unlynx} for more details. The choice of 
parameters depends on the application's privacy policy and is out of the scope of this paper.

\begin{protocol} {Collective Differential Privacy (CDP)}
	\addtocounter{protocol}{-1}
	\refstepcounter{protocol}\label{CDP}
	\vspace{-0.5em}
	\small
	\textit{Input.} $\epsilon$ (defined in Section \ref{sub_ch:diffP}), $\Delta f$: query sensitivity, and $\theta$: quanta\\
	\small
	\textit{Output.} E$_K(\hat{n}_1, ..., \hat{n}_{\tilde{l}}$)\\
	\vspace{-0.5em}
	\small
	\textit{Initialization}
	\begin{enumerate}[1.]
		\small
		\setlength\itemsep{0em}
		\vspace{-0.4em}
		\item The distribution $LD = Laplace(0,\ \Delta f/\epsilon)$ is publicly agreed on.
		\item $LD$ is publicly sampled, using the quanta $\theta$, to a list of $\tilde{l}$ noise values $\tilde{n_1}, ..., \tilde{n}_{\tilde{l}}$.
	\end{enumerate}
\small
	\textit{Protocol.}
	\begin{enumerate}[1.]
		\small
		\setlength\itemsep{0em}
		\vspace{-0.2em}
		\item Each $CN$ privately and sequentially shuffles $\tilde{n_1}, ..., \tilde{n}_{\tilde{l}}$, producing E$_K(\hat{n_1}, ..., \hat{n}_{\tilde{l}}$).
		\item First elements of E$_K(\hat{n_1}, ..., \hat{n}_{\tilde{l}}$) are used as oblivious noise values and added to the query result.
		%\vspace{-1.5em}
	\end{enumerate}
	%\vspace{-1.0em}
\end{protocol}

\emph{Security Arguments.} We observe that the list of noise values is verifiably generated from the differential privacy parameters and that all the $CNs$ privately shuffle the values. This protocol's security is analyzed in details in UnLynx \cite{froelicher2017unlynx}.
\vspace{0.5em}
\subsubsection{\bf Query Execution Correctness}\label{subsec:queryCorrect}
We first describe how \sys provides auditability by enabling an efficient verification of the query execution correctness. The latter is achieved by guaranteeing results robustness and computation correctness. The first is ensured by limiting the $DPs$' values to be in a specific range (by means of range proofs) and the second by using ZKPs for all the $CNs$ computations.
\vspace{0.3em}
\paragraph{\bf Auditability}
To provide an efficient solution for the query verification, \sys relies on a set of $VNs$ that verify the query correctness in parallel to its execution and without affecting its runtime. After each operation, $Q$, the $CNs$ and $DPs$ create proofs of correct computations or value range that they sign with their private key (to provide authentication). Their signed proofs are sent to all the $VNs$. This enables an efficient query execution as the proof creation and verification are executed independently from it.

\begin{figure}[!h]
	%\vspace{-1.0em}
	\centering
	\includegraphics[width=0.7\columnwidth]
	{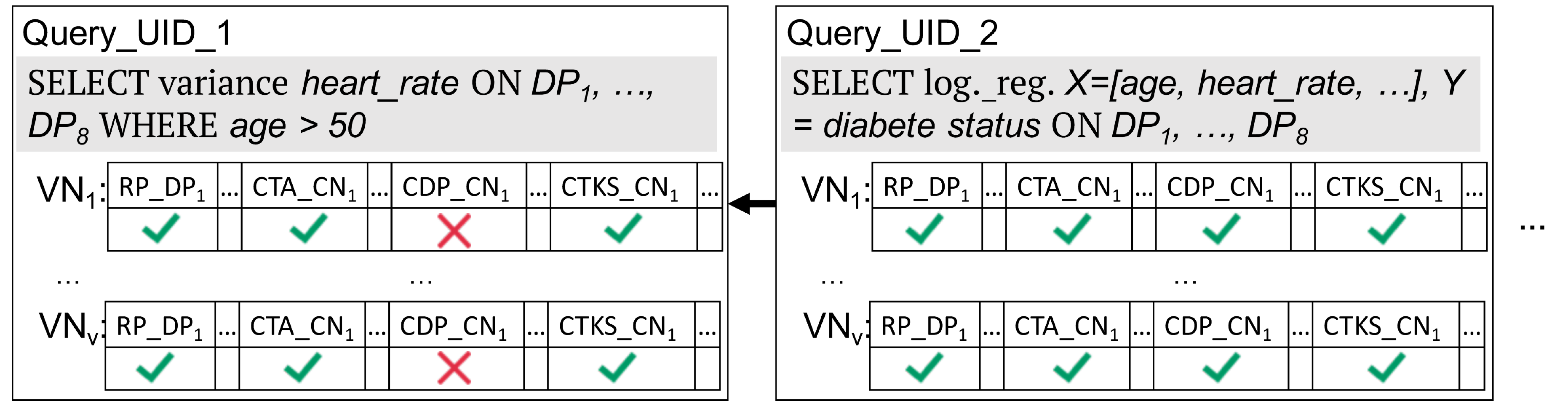}
	%\vspace{-1.5em}
	\caption{\textit{Proof blockchain}. Each block contains Query ID and content, and each $VN$'s \textit{query-proofs map}. RP stands for range proof.}
	\label{fig:blockchain}
	\vspace{-0.5em}
\end{figure}

\david{In order to implement this solution, we can rely on the distributed architecture of the $VNs$ and can provide integrity and immutability by using a blockchain, i.e., the \textit{proof blockchain}. This enables the public and immutable storage of both the query and its verification results.} Moreover, it enables an efficient and lightweight verification of the query correctness. An auditor, e.g., $Q$, has only to request the block corresponding to the query, to verify the $VNs$ signatures and to check the query verification results. We detail this in Protocol \ref{queryVerification} and show an example of the \textit{proof blockchain} in Figure \ref{fig:blockchain}.

\begin{protocol} {Query Verification}
	\addtocounter{protocol}{-1}
	\refstepcounter{protocol}\label{queryVerification}
	\vspace{-0.5em}
	\small
	\textbf{Query}\\
	{$Q$:}
	\vspace{-0.5em}
	\begin{enumerate}[1.]
		\small
		\item $Q$ signs and broadcasts the query to the $VNs$.
	\end{enumerate}
\small
	{$VNs$:}
	%\vspace{-0.5em}
	\begin{enumerate}[1.]
		\small
		\item Each $VN$ verifies $Q$'s signature.
		\item Each $VN$ deterministically derives the list of expected proofs for the query. It initializes a \textit{query-proofs map} that stores the result of the verification for each proof: \texttt{true}, \texttt{false},  \texttt{not received} (before a predefined timeout).
	\end{enumerate}
\small
	\textbf{Query Execution.}\\
	\small
	\textit{$DP$ or $CN$:}
	%\vspace{-0.5em}
	\begin{enumerate}[1.]
		\setlength\itemsep{0em}
		\small
		\item A $DP$ or $CN$ executes an operation,  then creates, signs and sends the corresponding proof to the $VNs$.
	\end{enumerate}
\small
	{$VNs$:}
	%\vspace{-0.5em}
	\begin{enumerate}[1.]
		\small
		\setlength\itemsep{0em}
		\item Each $VN$ verifies the prover's signature.
		\item Each $VN$ verifies the proof and stores the result in its \textit{query-proofs map}.
		\item Each $VN$ stores the proof in its local (key, value)-database. The key is uniquely and deterministically derived from the query, the prover's ID and the proof type.
	\end{enumerate}
\small
	\textbf{End of Query Execution (or timeout).}\\
	\small
	{$VNs$:}
	%\vspace{-0.5em}
	\begin{enumerate}[1.]
		\small
		\setlength\itemsep{0em}
		\item One of the $VNs$ (e.g., chosen in a round-robin fashion) gathers all $VNs$' \textit{query-proofs maps}.
		\item The same $VN$ creates a block containing the \textit{Query Unique ID}, the \textit{Query} and all the \textit{query-proofs maps}.
		\item The block is sent around such that each $VN$ checks that its \textit{query-proofs map} and the query are correctly saved. If this is the case, the $VN$ signs the block.
		\item The $VNs$ run a consensus algorithm such that a block signed by a threshold $f_h$ of $VNs$ is consistently added to the blockchain. Each $VN$ keeps a local copy of the blockchain.
		%\vspace{-1.5em}
	\end{enumerate}
	%\vspace{-1.5em}
\end{protocol}
%\vspace{-2.5em}
\emph{Security Arguments.} If an entity trusts a threshold $f_h$ of the $VNs$, it can verify the query correct execution by checking the corresponding block in the \textit{proof blockchain}. The verifier can check that $f_h$ nodes agree on the correctness of the proofs. A block is created for every query, even if the proofs are wrong, thus enabling any entity to determine which parties were involved in incorrectly computed queries. Otherwise, as all the proofs are universally verifiable and stored by all $VNs$, an auditor, not trusting $f_h$ of the $VNs$, can request the proofs from a subset of them and check the proofs by itself.
\vspace{0.5em}
\paragraph{\bf Results Robustness}
If \david{the querier defines a query with range boundaries on the $DPs$' values}, the $DPs$ are requested to create proofs of range by following the algorithm detailed in Algorithm \ref{alg:the_alg}. This algorithm is built by adapting the $[0,u^l)$-range proof scheme proposed by Camenisch et al. \cite{camenisch2008efficient} to the Anytrust model. In this algorithm, the prover, i.e., $DP$, writes its secret value $m$ in base-$u$ and commits to the $u$-ary digits by using the $CN_i$s' signatures on these digits ($A_{i,b}$ in Algorithm \ref{alg:the_alg}). The $l$ created commitments complete the proof. To adapt this algorithm to the Anytrust model, the $DP$ must compute multiple proof elements, i.e., $c$, $V_{i,j}$, $a_{i,j}$, by combining all $CNs$' signatures, i.e., $Z_i$, $A_{i,b}$. This ensures that the $DP$ uses at least one $CN$'s signature for which it does not know the underlying secret.  The same transformation in \cite{camenisch2008efficient} can be applied to generalize the proof to any range $[b_l,b_u)$. 

\emph{Security Arguments.} Both the correctness and the zero-knowledge property of the range proof are proven by Camenisch et al. \cite{camenisch2008efficient}.
\begin{algorithm}
	\caption{Input Range Validation in Anytrust Model}
	\label{alg:the_alg}
	\renewcommand{\thealgorithm}{}
	\small
	A $DP$ proves that its secret $m \in [0,u^l)$, where $u$ and $l$ are two integers. $C_2= mB + r\Omega$ corresponds to the right part of E$_\Omega$($m$)$=(C_1,C_2)$. $e()$ is a pairing function (bilinear map \cite{camenisch2008efficient}) on an Elliptic Curve and $H$ is a hash function.
	
	\small
	\noindent \textbf{Initialization:}
	\begin{algorithmic}[1]
		\small
		\STATE Each $CN_i$ picks a random $x_i \in \mathbb{Z}_p$ and computes $Z_i \gets Bx_i$, $A_{i,b} \gets B(x_i + b)^{-1} $ $\forall b \in \mathbb{Z}_u$.\\
		\STATE All $Z_i$ and $A_{i,b}$ are made public.\\
	\end{algorithmic}
	
	\small
	\textbf{Proof Creation:}
	\begin{algorithmic}[1]
		\small
		\STATE $DP$ computes value $c = H(B,C_2,\textstyle\sum_i{Z_i})$ and\\ %JR OK: Is H defined here?
		\FOR{each $j \in \mathbb{Z}_l $ such that $ m = \textstyle\sum_{j}{m_j u^j}$} 
		\STATE Pick three uniformly-random values $s_j,t_j,v_j \in \mathbb{Z}_p $ %JR OK: uniformly random?
		\FOR {each computing node $CN_i$} %JR OK: define S (avoid uppercase for an integer)
		\STATE  $V_{i,j} = A_{i,m_j}v_j$
		\STATE $a_{i,j} \gets -s_j \cdot e(V_{i,j},B) + t_j \cdot e(B,B)$
		\ENDFOR
		\STATE $z_{v_j} \gets t_j-v_j c\ (mod\ p)$ and $z_{m_j} \gets s_j-m_j c\ (mod\ p)$
		\ENDFOR
		\STATE $DP$ picks $n \in \mathbb{Z}_p$ and computes $z_r = n-rc\ (mod\ p)$ and $D \gets \textstyle\sum_{j}{Bu^j s_j} + \Omega n$\\
		\STATE $DP$ publishes $proof = \{C_2, c, z_r, z_{v_j}, z_{m_j}, D, a_{i,j}, V_{i,j}\} $ $\forall j \in \mathbb{Z}_l$ and $\forall i \in \{1,...,\#CN\}$.\\
	\end{algorithmic}
	
	\small
	\textbf{Proof Verification:}
	\begin{algorithmic}[1]
		\small
		\STATE Any entity can check that:\\
		$D = C_2c + \Omega z_r + \textstyle\sum_{j}{Bu^j z_{m_j}} $ and
		$a_{i,j} = e(V_{i,j},Z_i)c -z_{m_j} \cdot e(V_{i,j},B) + z_{v_j} \cdot e(B,B)$, $\forall j \in \mathbb{Z}_l$ and $\forall i \in \{1,...,\#CN\}$.
	\end{algorithmic}
	\vspace{-0.5em}
\end{algorithm}
%\vspace*{-1.2em}

 \noindent These proofs are universally verifiable and sound in the Anytrust model. The latter comes from the fact that the elements depending on the $CNs$' secrets $x_i$ are computed as a combination of all their public signatures. As at least one $CN_i$ is honest-but-curious, one of the $x_i$ is unknown (not revealed) to the $DP$ (prover).

\vspace{0.2em}
\paragraph{\bf Computation Correctness}
In order to ensure the correctness of the query execution, each computation executed by a $CN$ has to be proven correct.
\vspace{-0.2em}
\begin{itemize}[leftmargin=*]
	\setlength\itemsep{0em}
	%\item \emph{Encoding \& Encryption}. The $DPs$ compute range proofs for their values. We discuss
	\item \emph{Collective Tree Aggregation}. The $CNs$ provide to-be-aggregated input ciphertexts and the resulting ciphertexts that constitute the ZKP.
	\item \emph{Collective Tree Obfuscation}. The $CNs$ produce an obfuscation proof by relying on Expression \eqref{eq:proof_equation} in Section \ref{sub_ch:zkp_proofs}. Each $CN_i$ multiplies $C$ by $s_i$ to obtain the obfuscated ciphertext $(C'_1,C'_2)$ with (a) $C'_1 = s_iC_1$ and (b) $C'_2 = s_iC_2$. For both equations, $y_1=s_i$ is the discrete logarithm; we have the public values $A=C'_1$, $A_1=C_1$ for (a) and $A=C'_2$, $A_1=C_2$ for (b), which constitute the proof.
	\item  \emph{Collective Differential Privacy}. In this protocol, each $CN$ sequentially executes a Neff shuffle and produces the corresponding ZKP of correctness described in Section \ref{sub_ch:verifiable_shuffle}. This proof basically contains the input and output lists, the public key encrypting the ciphertexts, and commitment values.
	\item \emph{Collective Tree Key Switching}. The $CNs$ create the ZKP by applying Equation \eqref{eq:proof_equation} in Section \ref{sub_ch:zkp_proofs}, in which we have $y_1=k_i$, $y_2 = \alpha_i$, the discrete logarithms of $k_iB=K_i$ and $\alpha_iB$, respectively. All points $K_i$, $\alpha_iB$, $A=w_{i,2}$, $A_1 = -rB$ and $A_2 = K'$ are made public and do not leak any information about the underlying secrets. %Again, the aggregations are proven with the ciphertexts.
\end{itemize}
\emph{Security Arguments.} We rely on proofs that are universally verifiable and zero-knowledge. They do not affect data confidentiality beyond what can be inferred from the proven facts themselves.
%\vspace{-0.7em}
\subsection{Drynx Optimized Design}\label{sec:performance} 
We present \sys's final query execution pipeline, before describing how the query verification's performance can be optimized.
\vspace{0.2em}
\subsubsection{\bf Full Query Execution Pipeline}\label{subsec:parallelization}
\begin{figure}[h!]
	\centering
	%\vspace{-0.5em}
	\includegraphics[width=0.6\columnwidth]
	{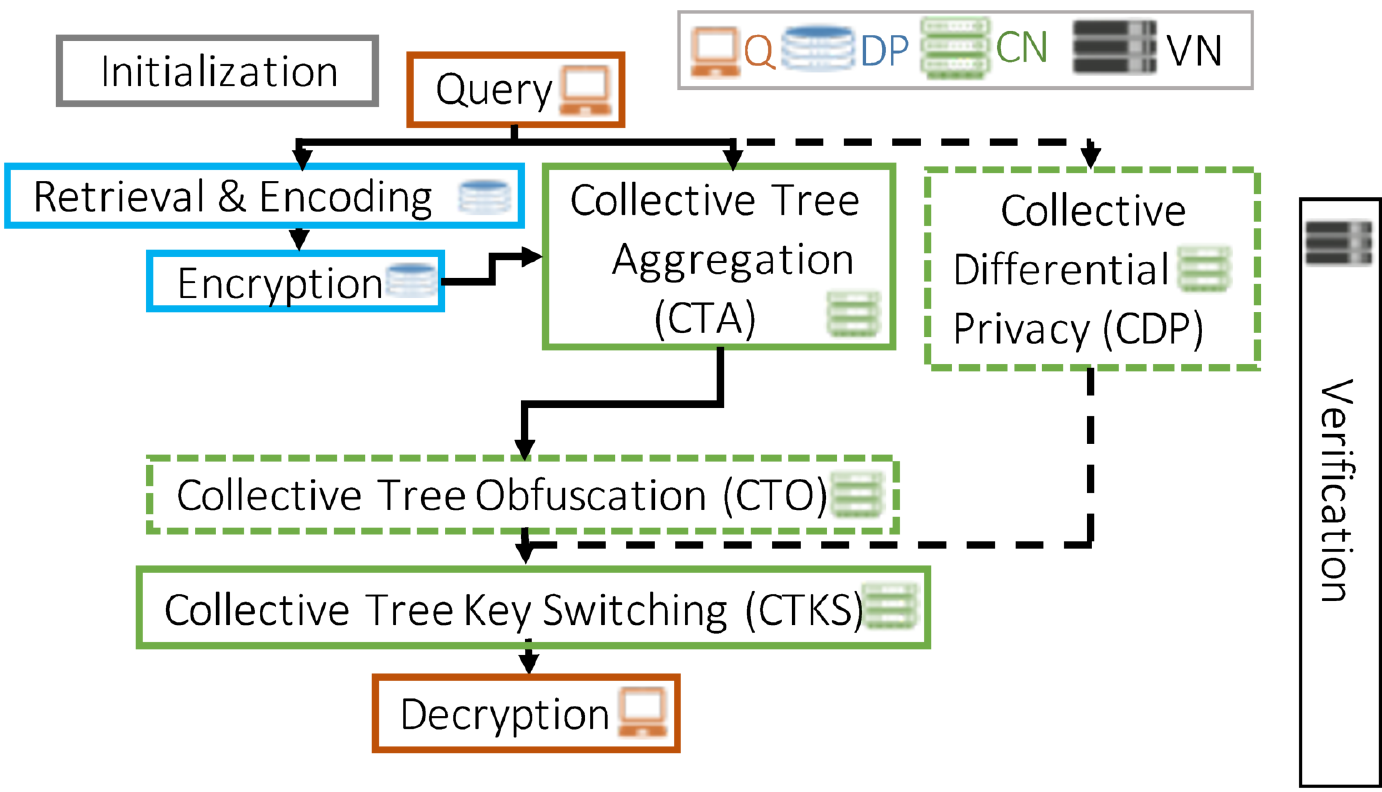}
	\vspace{-0.5em}
	\caption{\sys's complete optimized query-execution. Arrows represent causal links. Steps without direct links can be executed independently and dashed steps are optional.}
	\label{fig:drynx}
	%\vspace{-0.5em}
\end{figure}

We show \sys 's full pipeline in Figure \ref{fig:drynx}. Query execution and verification are executed concurrently and multiple steps of the query execution can be executed in parallel. The $CNs$ aggregate each $DP$'s response in $CTA$, as soon as they receive it. The noise generated from the $CDP$ has to be added after all the results have been aggregated. However, if the differential privacy parameters are predefined, this protocol can be executed independently from the other steps or even pre-computed. 

\subsubsection{\bf Probabilistic Query Verification}\label{subsec:probaVerif}
To improve the performance of the query verification, we enable a probabilistic verification of the proofs by the $VNs$. \david{We show that this strategy still enables a verifier to detect a misbehaving entity with a high probability, yet considerably improves performance (see Section \ref{sec:eval}).} A \emph{proof} for a specific operation (e.g., $CTKS$ for a set of ciphertexts \david{$S$}) can have multiple \textit{sub-proofs} (e.g., $CTKS$ for one ciphertext \david{$C \in S$}). One \textit{proof} is considered incorrect if one or more of the \textit{sub-proofs} is incorrect. We introduce the two thresholds $T$ and $T_{sub}$ that define the probability of verifying a single \textit{proof} and a \textit{sub-proof}, respectively. We modify the $VNs$' operations in step 2 of the \textit{Query Execution} described in Protocol \ref{queryVerification}, by adding this probabilistic verification based on $T$ and $T_{sub}$. \david{Each $VN$ stores all the \textit{proof} it receives. It then generates a random value $r \in [0,1]$; if $r<T$, it starts the probabilistic verification of the \textit{sub-proofs}. For each \textit{sub-proof}, the same method is applied, using $T_{sub}$.} 

\emph{Security Arguments.} The probabilistic verification does not necessarily compromise the security level of the system, given that the verification of each proof is redundantly done by each $VN$. A \textit{proof} is verified with a probability $p_{ver} = 1-(1-T)^{N_{VN}}$, where $N_{VN}$ is the number of $VNs$, and a \textit{sub-proof} with a probability $p_{ver_{sub}} = 1- ((1-T)+T(1-T_{sub}))^{N_{VN}}$. \david{The probability that a \textit{proof} or a \textit{sub-proof} is verified by at least $f_h$ nodes is}
%This ensures that a \textit{proof} or \textit{sub-proof} is verified by $f_h$ nodes with a probability:
%\vspace{-0.2em} 
\[
P_{f_{h}} = \textstyle\sum\limits_{i=f_h}^{N_{VN}}\binom{N_{VN}}{i}p^i(1-p)^{N_{VN}-i},
%P_{maj_{ver}} = \textstyle\sum\limits_{i=1+\lfloor(\frac{N_{VN}}{2})\rfloor}^{N_{VN}}\binom{N_{VN}}{i}p^i(1-p)^{N_{VN}-i},
%\vspace{-0.3em}
\]
where $p$ is either $p_{ver}$ \david{(for a \textit{proof})} or $p_{ver_{sub}}$ \david{(for a \textit{sub-proof})}. For example, if $N_{VN}=7$, $T=1$ and $T_{sub}=0.3$, all the \textit{proofs} are at least partially verified and each \textit{sub-proof} is verified by $f_h=5$ $VNs$ with $P_{f_{h}}=98.48\%$. Each \textit{sub-proof} is thus verified by at least $f_h$ of the $VNs$ with a high probability. Due to the honesty assumption, a \textit{sub-proof} is at least verified by one honest $VN$ with a high probability. \david{Moreover, the thresholds $T$ and $T_{sub}$ can be set to arbitrarily reduce the probability that one \textit{sub-proof} is not verified by at least one honest node.} Therefore, if all the $VNs$ that participated in the verification agree on the result, the auditor knows the \textit{proof} is correct, otherwise it can either choose to only trust some of the $VNs$ or fetch all proofs and verify them itself, \david{as all the proofs are universally verifiable. For example, an auditor can choose to verify only the proofs that were not checked by any of the $VNs$ she trusts.}

%% file: securityAnalysis.tex
\vspace{-0.5em}
\section{Security Analysis}\label{sec:securityAnalysis}
We employed only existing, peer-reviewed cryptographic schemes and discussed the composability of the security of the different blocks in previous sections. We corroborate these arguments with a brief summary of the security analysis.
\begin{itemize}[leftmargin=*]
	\setlength\itemsep{0em}
\item \emph{Data confidentiality}. In Section \ref{subsec:dataconf}, we sketched the proof for confidentiality in our simplified system and discussed in Section \ref{sec:security} how further design choices do not hinder confidentiality. In summary, data confidentiality is ensured as the data are always encrypted and no operation, e.g., $ZKP$ creation, affects it.
\item \emph{$DPs$' privacy}. $DPs$ can privately decide whether to answer a query, and differential privacy is ensured for the $DPs$ and individuals, which protects them from potential inferences stemming from the release of end results. The latter is ensured  in \sys by blindly adding noise, sampled from a specific distribution, to the query end results. As described in Section \ref{subsec:dpPrivacy}, this noise can be verified to be from a specific distribution (e.g., Laplacian) and no entity knows which noise value is added.
\item \emph{Results \robust}. This is ensured as all $DPs$' values can be verified to be within a certain range and all $CNs$' computations must be proven correct, as depicted in Section \ref{subsec:queryCorrect}. By enforcing the generation of range proofs by $DPs$, we protect against strong outliers, maliciously or erroneously input, which can significantly distort the query results. $DPs$ can still input incorrect values, but their influence on the final result is limited. We give an intuition on how robust a computation is against such behavior in Section \ref{prototype_evaluation}.  
\item \emph{Computation correctness.} The proofs of correct computations (Section \ref{subsec:queryCorrect}) ensure that the $DPs$' answers are correctly aggregated ($CTA$) and that the remaining steps ($CTO$, $CTKS$, $CDP$) are correctly executed.
\end{itemize}

%% file: encoding.tex
%\vspace{-1.0em}
\section{Encodings}\label{sec:encoding}
We present a set of statistical computations that can be executed in \sys. We then explain how to instantiate \emph{encodings} (Definition 1) for the training of both linear and logistic regression machine-learning models. We adapt the logistic regression solution, proposed by Aono et al. \cite{aono2016scalable}, to our framework, thus enabling $Q$ to train this model in a verifiable and privacy-preserving way, even in the presence of a strong adversary. Some of the encodings are adapted from the Corrigan-Gibbs and Boneh \cite{corrigan2017prio} system and improved upon. 

\textbf{Numerical Statistics.} Table \ref{tab:encodings} lists a set of simple statistics that can be performed with \sys. The \texttt{sum}, \texttt{mean}, \texttt{variance}, \texttt{std. deviation}, \texttt{cosine similarity (cosim)} and \texttt{R}$^2$ operations are executed by requiring the $DPs$ to send the result of their local and partial statistic computation. As an example, for \texttt{variance}, each $DP_i$ locally computes the sum of the values (records) $h_j$ that match the query, ($\scriptstyle\sum_{j=1}^{c_i} h_{j}$) where $c_i$ is $DP_i$'s dataset cardinality, the square of those same values ($\scriptstyle\sum_{j=1}^{c_i} h_{j}^2$) and generates $\rho(\bar{r_i}) = \scriptstyle ([\sum_{j=1}^{c_i}h_j,\scriptstyle\sum_{j=1}^{c_i} h_{j}^2],c_i)$. These values are independently aggregated among all $DPs$ and the overall variance is computed by $Q$, after decryption, using the corresponding $\pi$ (defined in Table \ref{tab:encodings}). For the \texttt{frequency count}, $DPs$ are expected to send the vector $\mathbf{V_i}$ filled with the number of occurrences ($fc$) for specific values. The \texttt{cosine similarity} is computed between two vectors $\phi$ and $\boldsymbol{\bar{\phi}}$, where each $DP_i$ holds a subset of the coefficients of each vector.

\begin{table}[!h]
	%\vspace{-0.8em}
	\centering
	\small
	\begin{tabular}{| c | c | c |}
		\hline
		Operat. ($f$) &  $\pi$ (on $N$ $DPs$) & $\rho$ \\
		                     &                                      & ($\scriptstyle{\mathbf{V_i}} = [v_{i,1}, ..., v_{i,d}], \scriptstyle c_i$)\\ \hline
		sum  & $\scriptstyle\sum_{i=1}^{N} v_{i,1}$ & ([$ \scriptstyle\sum_{j=1}^{c_i} h_{j}$], $\scriptstyle c_i$) \\   \hline
		mean  & $\frac{\scriptstyle\sum_{i=1}^{N} v_{i,1}}{\scriptstyle\sum_{i=1}^{N} c_{i}}$ & ([$ \scriptstyle\sum_{j=1}^{c_i} h_{j}$], $\scriptstyle c_i$) \\   \hline
		variance  & $\sigma^2 = \frac{\scriptstyle\sum_{i=1}^{N} v_{i,2}}{\scriptstyle\sum_{i=1}^{N} c_{i}} - (\frac{\scriptstyle\sum_{i=1}^{N} v_{i,1}}{\scriptstyle\sum_{i=1}^{N} c_{i}})^2$ & ([$ \scriptstyle\sum_{j=1}^{c_i} h_{j}$, $ \scriptstyle\sum_{j=1}^{c_i} h_{j}^2$],  \\ 
		std. dev. & $\scriptstyle \sigma = \sqrt{\sigma^2}$ & $\scriptstyle c_i$) \\ \hline
		$\scriptstyle{AND/OR}$ & $\scriptstyle\sum_{i=1}^{N} v_{i,1} \stackrel{?}{=} 0$ & ([$\scriptstyle R_{j}$],$\scriptstyle c_i$) or ([$\scriptstyle b_{j}$], $ \scriptstyle c_i$) \\ \hline
		min/max & $l/rm_{\neq 0}(\scriptstyle\sum_{i=1}^{N} v_{i,1}, ..., $ & ([$\scriptstyle R_{j,1}$, ..., $\scriptstyle R_{j,d}$], $\scriptstyle c_i$) \\ 
		& $\scriptstyle\sum_{i=1}^{N} v_{i,d})$ &or ([$\scriptstyle b_{j,1}$, ..., $\scriptstyle b_{j,d}$], $\scriptstyle c_i$) \\ \hline
		frequ. count & $\scriptstyle\sum_{i=1}^{N} v_{i,1}, ..., \scriptstyle\sum_{i=1}^{N} v_{i,d}$ & ([$\scriptstyle fc_{j,1}$, ..., $\scriptstyle fc_{j,d}$], $\scriptstyle c_i$) \\ \hline
		set int/un & $\scriptstyle\sum_{i=1}^{N} v_{i,1}, ..., \scriptstyle\sum_{i=1}^{N} v_{i,d}$ & ([$\scriptstyle R_{j,1}$, ..., $\scriptstyle R_{j,d}$], $\scriptstyle c_i$) \\ 
		&  &or ([$\scriptstyle b_{j,1}$, ..., $\scriptstyle b_{j,d}$], $\scriptstyle c_i$) \\ \hline
		cosim & $\scriptstyle s(\phi,\overline{\phi}) = \scriptstyle\frac{\scriptstyle\sum_{i=1}^{N} v_{i,1}}{\sqrt{\scriptstyle\sum_{i=1}^{N}v_{i,2}} \sqrt{\scriptstyle\sum_{i=1}^{N}v_{i,3}}}$ & ([$ \scriptstyle\sum_{j=1}^{c_i} \phi_{j}\overline{\phi}_{j}, \scriptstyle\sum_{j=1}^{c_i} \phi_{j}^2,$ \\ 
		& & $ \scriptstyle\sum_{j=1}^{c_i} \overline{\phi}_{j}^2$], $\scriptstyle c_i$) \\ \hline 
		$R^2$ & $ 1 - \frac{\scriptstyle\sum_{i=1}^{N} v_{i,3}}{\sigma^2}$ & ([$ \scriptstyle\sum_{j=1}^{c_i} y_{j}$, $ \scriptstyle\sum_{j=1}^{c_i} y_{j}^2$ \\
		& &  $\scriptstyle\sum_{j=1}^{c_{i}} (y_j - \hat{y_j})^2$], $\scriptstyle c_i$) \\ \hline
	\end{tabular}
	\vspace{-0.5em}
	\caption{Example set of \enco instantiations. All $DPs$ \enco{}s ($\rho$) are then aggregated such that $Q$ computes $\pi$ at the end.}
	\label{tab:encodings}
	%\vspace{-0.9em}
\end{table}

\textbf{Bit-Wise Statistics.} As depicted in Table \ref{tab:encodings}, bit-wise operations can be executed in two ways: Each $DP_i$ either (1) sends a random encrypted integer $R$ or (2) sends an encrypted bit $b$. For (1), in the \texttt{OR} (resp. \texttt{AND}) case, each $DP_i$ is requested to send an encrypted integer $E_K(R_i)$, where $R_i = 0$ if the input is $0$ (resp. $1$), and a random positive integer otherwise. The \texttt{OR} (resp. \texttt{AND}) expression is $true$ (resp. $false$) if the sum $\textstyle\sum R_i > 0$. $Q$ obtains the final result by testing if the output is 0 or not. The result of this operation can be erroneous if $\textstyle\sum R_i \equiv 0\ mod (\#G)$, or in other words, if the order $\#G$ of the Elliptic Curve subgroup divides the sum of all $DPs$' random values. This happens only with a probability smaller than ${1}/{(\#G-1)}$ (proof in Appendix \ref{error_proba}). This probability is close to 0 as $\#G$ is much bigger than the decryptable plaintext values, and can be further reduced by repeating the query. Alternatively, in (2) each $DP_i$ has to send $b_{i,j} = 0$ or $b_{i,j} = 1$ encrypted value. This eliminates the error probability but requires more computations and proofs of correctness, as the $DPs$ have to prove that their values are in $\{0,1\}$, and a $CTO$ protocol (Section \ref{par:bitwise}) has to be executed to preserve privacy.
The \texttt{min} (resp. \texttt{max}) is computed by applying the \texttt{or} operation element-wise among vectors $\mathbf{V_i}$. Each $DP_i$ computes its local min (resp.  \texttt{max}) $m_{DP_i}$ in a specified range, e.g., [0:100], which is represented by $\mathbf{V_i}=[b_{i,0}, ...,b_{i,100}]$. Each $b_{i, j} >  m_{DP_i}$ (resp. $b_{i, j} <  m_{DP_i}$)  is encoded with a `1' (or random) and a `0' otherwise. The \texttt{min} (resp. \texttt{max}) across all $DPs$ corresponds to the leftmost (resp. rightmost) position with a `1' in the vector resulting from the \texttt{OR} operation. Similarly, the \texttt{set intersection} (\texttt{resp. union}) is computed by using the \texttt{AND} (resp. \texttt{OR}) operation element-wise on the vectors $\mathbf{V_i}$.

\textbf{Regression Models.} 

\noindent \emph{Linear Regressions}. We assume a dataset distributed over the $DPs$ with $D$ features $x_1, ..., x_D$ and a label value $y$ such that $y \approx c_0 + c_1 \times x_1 + c_2 \times x_2 + ... + c_D \times x_D$. \sys computes the least-squares linear fit over all the $DPs$ by building a system of $D+1$ equations that $Q$ can use in order to compute the linear regression coefficients $c_0, c_1, c_2, ..., c_D$:
\begin{gather}
\small
\setlength\arraycolsep{2.5pt}
\begin{pmatrix} n & \scriptstyle\sum x_{\mu,1} & ... & \scriptstyle\sum x_{\mu,D}\\
\scriptstyle\sum x_{\mu,1} & \scriptstyle\sum x_{\mu,1}^2 & ... & \scriptstyle\sum x_{\mu,1}x_{\mu,D}\\
... & ... & ... & ... \\
\scriptstyle\sum x_{\mu,D} & \scriptstyle\sum x_{\mu,1}x_{\mu,D} & ... & \scriptstyle\sum x_{\mu,D}^2\\
\end{pmatrix}
\setlength\arraycolsep{1.5pt}
\begin{pmatrix} c_0 \\ c_1 \\ ... \\ c_D
\end{pmatrix}
\approx
\setlength\arraycolsep{1.5pt}
\begin{pmatrix}
\scriptstyle\sum y_\mu \\ \scriptstyle\sum y_\mu{}x_{\mu,1} \\ ... \\ \scriptstyle\sum y_\mu{}x_{\mu,D}
\end{pmatrix}
\end{gather}
where all the sums are between $\textstyle \mu=1$ and $\textstyle \mu=\textstyle\sum_{i=1}^{N}c_i$. Each $DP_i$ sends $\textstyle\sum_{j=1}^{c_i} x_{j,\eta}$, $\textstyle\sum_{j=1}^{c_i} x_{j,\eta}x_{j,\zeta}$, $\textstyle\sum_{j=1}^{c_i} y_j$, $\textstyle\sum_{j=1}^{c_i} y_jx_{j,\eta}$, $\forall \eta,\zeta \in \{1, 2, ..., D\}$, $\eta \neq \zeta$.

\noindent \emph{Logistic Regressions}. 
We consider again a dataset of $N_{da}$ records (distributed among the $DPs$) with a dimension $D$ where each record $x^{(\mu)} = (1, x^{(\mu)}_1, \cdots, x^{(\mu)}_D) \in R^D$ consists of $D$ features and an offset term of 1, and is associated with a label $y^{(\mu)} \in \{0,1\}$. The original logistic regression cost function is
%\vspace{-0.3em}
\begin{align*}
\scriptstyle J(\theta) = \frac{1}{N_{da}} \sum_{\mu=1}^{N_{da}} \Big[ -y^{( \mu)} \log(h_{\theta}(x^{(\mu)})) - (1 - y^{(\mu)}) \log(1 - h_{\theta}(x^{(\mu)})) \Big] + lr_\theta,
\end{align*}
%\vspace{-0.2em}
where $h_{\theta}(x)={1}/{(1 + \exp(\scriptstyle\sum_{\eta=0}^D \theta_\eta x_\eta))}$ and $lr_\theta = \frac{\lambda}{2N_{da}} \scriptstyle\sum_{\eta=1}^{D} \theta_\eta^2$, $\lambda$ is the L2-regularization parameter.
$J(\theta)$ can be approximated by a linear function
%\vspace{-0.1em}
\begin{align*}
\scriptstyle J_{a}(\theta) = \Big[\frac{1}{N_{da}} \scriptstyle\sum_{\tau=1}^k \scriptstyle\sum_{r_1, ..., r_\tau = 0}^D a_\tau (\theta_{r_\tau} \cdots \theta_{r_\tau}) A_{\tau, r_1, ..., r_\tau} - a_0 \Big] + LR_\theta,
%\vspace{-0.5em}
\end{align*}
by using the fact that $\scriptstyle \log ( \frac{1}{1+\exp(x)}) \approx \scriptstyle\sum_{\tau{}=0}^k a_\tau x^\tau$, where $a_0, a_1, ...,$ $a_k$ can be chosen as the $k+1$ first coefficients of the Taylor expansion of $\scriptstyle \log ( \frac{1}{1 + \exp(x)} )$, or as the coefficients of the quadratic approximation that minimizes the area between the original function and its approximation. The $A_{\tau, r_1, \cdots, r_\tau}$ coefficients are defined by
%\vspace{-0.3em}
$$\scriptstyle A_{\tau, r_1, \cdots, r_\tau} = \scriptstyle\sum_{\mu=1}^{N_{da}} a^{(\mu)}_{\tau, r_1, \cdots, r_\tau}  = \scriptstyle\sum_{\mu=1}^{N_{da}} ( y^{(\mu)} - y^{(\mu)}(-1)^\tau - 1) ( x^{(\mu)}_{r_1} \cdots x^{(\mu)}_{r_\tau} ), $$
where the $a^{(\mu)}_{\tau, r_1, \cdots, r_\tau}$  are computed and encrypted by the $DPs$ before being collectively aggregated by the $CNs$.

\textbf{Neutral Response.} A neutral response for \texttt{and} and \texttt{set intersection} is $O=[1, ..., 1]$, and $O=[0, ..., 0]$ for other operations.

\textbf{Optimized and Iterative Encoding} \sys{} can also be used in order to execute iterative processes, e.g., a k-means algorithm. In this case, each iteration can simply be mapped to a query sent to the system. An iterative process can also be used in order to optimize existing \enco{}s, such as the \texttt{min} and \texttt{max}. In their basic versions, these \enco{}s rely on a $d$-bit vector in which each bit represents a value in a predefined range of size $d=|b_u-b_l|$. This means that each $DP$ sends $d$ ciphertexts. This process can be optimized by using a binary-search iterative process as depicted in Protocol \ref{iterProcess}. In the \textit{Range Reduction} step, each query only requires one ciphertext per $DP$ and reduces by half the range of possible answers. This step is repeated until this range is reduced to a predefined size $EL$.
\david{It must be noted that the execution of other iterative processes would work in a similar way: For example, for a k-means algorithm \cite{hartigan1979algorithm}, $Q$ performs one iteration by executing one query that includes the centroids in clear; the $DPs$ then assign their points to the closest centroid before aggregating their points by cluster; then, the same operation is repeated among all $DPs$ by using \sys typical query workflow and $Q$ computes the new centroids. As in Protocol \ref{iterProcess} and as described below, this algorithm leaks the intermediate results. We do not address the problem of hiding the intermediate results, e.g,  by using differential privacy, in this work.}
%\vspace{-0.5em}
\begin{protocol}{Iterative Process (\texttt{max} example)}
	\small
	\setcounter{protocol}{\value{protocol}-1}
	\refstepcounter{protocol}\label{iterProcess}
	\vspace{-0.5em}
	\textit{Input.} Query = \texttt{max} in $ra = [b_l,b_u]$ and $EL$
	
	\small
	\textit{Output.} Max value
	
	\small
	\textit{Range Reduction:}
	\begin{spacing}{0.5}
	\begin{algorithmic}[1]
		\small
		\setlength\itemsep{0em}
		\WHILE{$|ra| > EL$}
		\STATE $Q$ sends  \texttt{SELECT OR ($\scriptsize \exists v \in [\lfloor\frac{(b_l+b_u)}{2}\rfloor,b_u]$) ON $\scriptstyle DP_1, .., DP_n$ }
		\IF{query returns $true$}
		\STATE $ra = [\lfloor\frac{(b_l+b_u)}{2}\rfloor,b_u]$
		\ELSE
		\STATE $ra = [b_l,\lceil\frac{(b_l+b_u)}{2}\rceil]$
		\ENDIF
		\ENDWHILE
	\end{algorithmic}
	\end{spacing}
\vspace{0.9em}
\small
	\textit{Final Step:}
	\begin{spacing}{0.5}
	\begin{algorithmic}[1]
		\small
		\STATE $Q$ sends \texttt{SELECT MAX $[b_l,b_u] $ ON $\scriptstyle DP_1,..., DP_n$ }
		\vspace{-0.2em}
	\end{algorithmic}
\end{spacing}
	\vspace{-1.2em}
\end{protocol}

\emph{Security Arguments.} For all \enco and in each query, $Q$ learns the elements of $\mathbf{V}$ (aggregated over all $DPs$) and the (approximate) number of samples considered $c$, as defined by \enco. 

For the iterative process, in the \textit{Range Reduction}, the $DPs$' answers remain confidential, but the range is sent in clear in each query thus revealed to other entities. $Q$ controls the size of the range of possible values that is leaked by defining an entropy limit $EL$. In the \textit{final step}, the \texttt{max} query is privately executed on the remaining range. This provides a tradeoff between performance and privacy (that we analyze in Section \ref{sec:eval}). The number of ciphertexts is lowered to $n =g + \lceil\frac{d}{2^g}\rceil,\ g=\lfloor log_2(\frac{d}{EL})\rfloor$, which reduces the amount of computations and proofs by a factor $\frac{d}{n}$. For example, if $Q$ wants to know the $DPs$' minimum value in $[0,1000)$ with $EL =100$, the workload is reduced by a factor of $7.8$ and the query leaks a range of 100 possible minimum values.

%% file: discussion.tex
\section{Discussion and Extensions}\label{sec:discussion}
We illustrate multiple extensions for \sys by relying on our use cases, $HDS$ and $PDS$ (Section \ref{sub_ch:running_ex}). %We discuss collusion resistance and availability in \sys before briefly describing how we envision authentication/authorization.

\textbf{Modularity.} \sys is highly modular and some of its security features can be enabled or disabled, depending on the application. For example, if results robustness is not required, input-range validation can be omitted without hindering \sys's execution and the remaining security guarantees are preserved. The same applies for $DPs$' privacy features, e.g., differential privacy.
	
For example, in $HDS$, each hospital (or $DP$) locally executes the query on multiple patient records and the range proofs can be omitted if the range of possible values is too broad or if the hospital is trusted to input correct values. Otherwise, the range boundaries have to be set accordingly. \david{In this case, the querier has to use her knowledge on the attributes involved (e.g., age is between 0 and 150) and the information she has on the $DPs$' data (e.g., $DPs$ have a maximum of $X$ data samples) to define the ranges. 
%As the query ranges are public, they can also be used to limit the $DPs$' operations, e.g., limit the number of records that can be included in the computation, thus limiting the influence of $DPs$ that have a high number of data records on the final result with respect to $DPs$ with less records, i.e, $DPs$ that have a lot of records have to use a sample of them.} 
In $PDS$, the ranges for the input values can be used to enforce tighter bounds (e.g., heart rate can only take values in [40,100] beats-per-minute) as each $DP$ has one data record. }
	
\sys also enables the collective protection of data at rest by having $DPs$ locally encrypt their data with the $CNs$' collective key $K$. This limits the flexibility of the system as $DPs$ are then required to pre-compute all necessary inputs (e.g., the square root of the values to enable the computation of the \texttt{variance}) and the range proofs before entering the encrypted data in their databases. It also requires a fixed set of $CNs$, as only they can operate with that pre-encrypted data.

\david{As mentioned before, \sys's primary goal is to guarantee $DPs$' privacy and still enable the queriers to obtain the results of computations performed over multiple databases. For this, \sys enables optional security and privacy features, such as differential privacy. These features can be enabled or disabled depending on the application requirements, hence enabling multiple trade-offs between security and privacy, performance and accuracy (see below).}
%These features can be enabled or disabled depending on the application requirements; e.g., if the application requires exact results and the participation of all $DPs$, then these features should not be used. We discuss the accuracy in \sys below. Finally, \sys can be used in multiple applications with heterogeneous requirements as it enables multiple tradeoffs between security and privacy guarantees, functionalities and performance.}
	
\textbf{Collusion Resistance.} Each participant can play multiple roles without hindering \sys's security. For example, in $HDS$, a hospital can be a $DP$ and also play the role of a $CN$, to ensure its data confidentiality without having to trust any other hospital. It can also be a $VN$ thus take part in the verification process.

\textbf{Availability.} \sys's privacy and security guarantees hold even in the case where multiple $CNs$ or $DPs$ become unavailable. \david{Any entity can leave or join the system without hindering \sys's operation, as long as they are not involved in a query under execution. In the event of a $CN$ becoming unresponsive during the query execution, the $CTA$ and $CTKS$ steps cannot be finalized, as they both require the participation of all $CNs$. Therefore, in this case, the process is stopped and $Q$ can request the same query by choosing another set of $CNs$, e.g., by excluding the faulty $CN$(s). An unresponsive $DP$ only reduces the number of responses included in the statistic being computed and does not disrupt \sys's process.} Standard mechanisms, e.g., limiting the rate at which queries are accepted, can be implemented in \sys to avoid DDoS attacks.

\david{\textbf{Accuracy.} There are several aspects that can influence output precision in \sys. (a) We first remark that the $DPs$' inputs to the system have to be approximated by fixed-point representation if they are floating values, as explained in Section \ref{EGHE}. \\
(b) \sys's encodings and query executions do not intrinsically hinder the accuracy of the computed results, as all operations are exact, as long as the target function is exactly \emph{encodable}. In fact, it is worth noting that the encoding for the logistic regression training is built from an approximation of the original cost function.\\
Additionally, (c) the $DPs$ can privately decide whether to answer a query; this choice can influence the final result. However, the number of samples considered in the computation, i.e., $c_i$ in Definition \ref{def:encoding}, is always sent to $Q$, who can then observe if this number changed since her last query. It also enables her to take an informed decision on the statistical significance of the results, to accept them or not. \\
(d) \sys can guarantee differential privacy by adding noise to the final result. In this case, \sys returns approximate results, and the accuracy loss depends on the chosen privacy parameters and the executed operation. The choice of these parameters and the perturbation introduced in the results 
is thus orthogonal to this work.\\ 
Finally, (e) malicious $DPs$ can try to distort the query result by inputting erroneous values. \sys limits malicious $DPs$' influence on the final result by enabling the querier to restrict the range of possible inputs. This bounds the perturbation that some $DPs$ can generate on the results. If the inputs were not bounded, one malicious $DP$ could completely distort the final result by inputting extreme values. It is difficult to provide hard numbers for the accuracy of \sys in the presence of malicious $DPs$, as it depends on many parameters such as the executed operation, the chosen input ranges, the number of $DPs$ and data records. Nonetheless, in Section \ref{sec:eval} we show how the use of ranges limits the influence of malicious $DPs$ in two examples.}
	%
%\david{In $PDS$, proofs of input range can be used to enforce tight bounds on the participants' inputs to the system, hence limit the influence of a malicious $DP$ on the final result. In $HDS$, each $DP$ computes on multiple patients' records and the ranges have to be set accordingly or can simply be omitted if the range of possible values is too broad.}
%
%\david{In $PDS$, a patient ($DP$) can have his data encrypted locally with $K$, thus protecting his data collectively, but limiting the flexibility of the system. In this case, the \enco and the input range validation proofs have to be pre-computed when the data are entered in the database. We notice that when it is not the case, \sys is highly dynamic as a subset of $CNs$ and the corresponding collective public key can be defined at each query.}
%
%\david{In $HDS$, we envision that an hospital can play multiple roles. It can participate as a $CN$ to ensure its data's confidentiality without having to trust any other hospital and/or as a $VN$ and thus take part in the verification process.}
%We remark that both differential privacy and input-range validation can be enforced or not. All \sys's parameters can be agreed on among the different parties or predefined in the system's initialization, depending on the application.

\textbf{Authentication/Authorization.} \david{Authentication and authorization fall out of the scope of this paper, but for the sake of completeness we briefly mention here that \sys can integrate off-the-shelf solutions based on federated or distributed architectures \cite{openid, oauth, kogias2016enhancing}.} 

%% file: evaluation.tex
\section{Performance Evaluation}\label{sec:eval}
We discuss our experimental setup and evaluate \sys's performance. We show that it scales almost (in some cases better than) linearly with the number of $CNs$, $VNs$ and $DPs$, and we compare \sys against existing solutions. We also discuss multiple security, privacy and performance tradeoffs.
\vspace{-0.6em}
\subsection{System Implementation}
We implemented \sys in Go \cite{golang}, and our full code is publicly available \cite{drynx}. We relied on Go's native crypto-library and on public advanced crypto-libraries \cite{dedis}. \david{For the implementation of the proofs' storage and verification, we use a skipchain \cite{nikitin2017chainiac}, which is made of blockchain-like blocks that, to enable clients to efficiently navigate arbitrarily on the chain, also contain back-and-forward pointers to older and future blocks.} We rely on a (private) permissioned blockchain \cite{danezis2015centrally}, as in our examples $HDS$ and $PDS$ (Section \ref{sub_ch:running_ex}), the participants, i.e., researchers, patients or hospitals, have to be known and authorized. However, \sys works independently of the blockchain type, and a permission-less blockchain can also be used in a less restrictive scenario. \sys works independently of the used Elliptic Curve; we tested it on the Ed25519 \cite{curve} and bn256 Elliptic Curves \cite{barreto2005pairing}. Both curves provide 128-bit security, and we used bn256 by default as it enables pairing operations (required for range proofs). Our prototype is built as a modular library of protocols that can be combined in multiple ways. The communication between different participants relies on TCP with authenticated channels (through TLS).
%\vspace{-1em}
\subsection{System Evaluation}\label{prototype_evaluation}
We used Mininet \cite{mininet} to simulate a realistic virtual network between the nodes; we restricted the bandwidth of all connections between nodes to 100Mbps and imposed a latency of 20ms on all communication links. We evenly distributed the $CNs$, $DPs$, $VNs$ and $Q$ on a set of 13 machines that have two Intel Xeon E5-2680 v3 CPUs with a 2.5GHz frequency that supports 24 threads on 12 cores and 256GB RAM.

\david{We begin our evaluation by studying how the different steps in \sys's pipeline can be executed in parallel. We then show that \sys's runtime only slightly increases when the number of records per $DP$ grows (and the number of $DPs$ remains constant).} %We show that \sys scales almost linearly with an increasing number of $DPs$, $CNs$ and $VNs$. 
%We observe how \sys's runtime changes with the computed statistic. We study the effects of the security tradeoffs on \sys's runtime and discuss its bandwidth and storage overheads. We give an intuition on results robustness in an example with a growing number of malicious $DPs$. We then compare, both qualitatively and quantitatively, \sys against the closest prior art solutions.

\david{In our \textbf{default setup}, we consider 6 \textbf{\textit{CNs}} and 7 \textbf{\textit{VNs}.}}
%and 60 \textbf{\textit{DPs}} (10 per \textbf{\textit{CN}})}. This corresponds to 60 hospitals sharing their data in $HDS$ (Section \ref{sub_ch:running_ex}); we exemplify \sys's use in $PDS$ by increasing the number of $DPs$ in multiple scenarios in the next section. 
We set the proof verification thresholds $T=1.0$ and $T_{sub}=0.3$  and show, in Section \ref{drynx_evaluation}, the effect of these thresholds on \sys's execution time. The joint use of these thresholds ensures that all the proofs are at least partially verified and that each \textit{sub-proof} is verified by $f_h$ $VNs$ with a probability of $98.5\%$. We show \sys'runtime without the $CDP$ protocol as $CDP$ can be pre-computed or run in parallel with other steps. We notice that the $CDP$'s runtime depends on the number of $CNs$ and on the size $\tilde{l}$ of the list of noise values. This creates a tradeoff between privacy and performance as a greater $\tilde{l}$ provides a higher privacy level, as it reduces $\delta= {1}/{\tilde{l}}$ but also increases the time to generate and shuffle the list of noise values. With a Laplacian distribution and $\tilde{l} = 100$, $CDP$'s runtime is 2.9 seconds with an overhead of 8.1 seconds for the proof verification. 

  \begin{figure*}[ht!]
	\centering
	\tiny
	\begin{subfigure}[t]{0.3\textwidth}
		\centering
		\includegraphics[width=1.0\columnwidth]
		{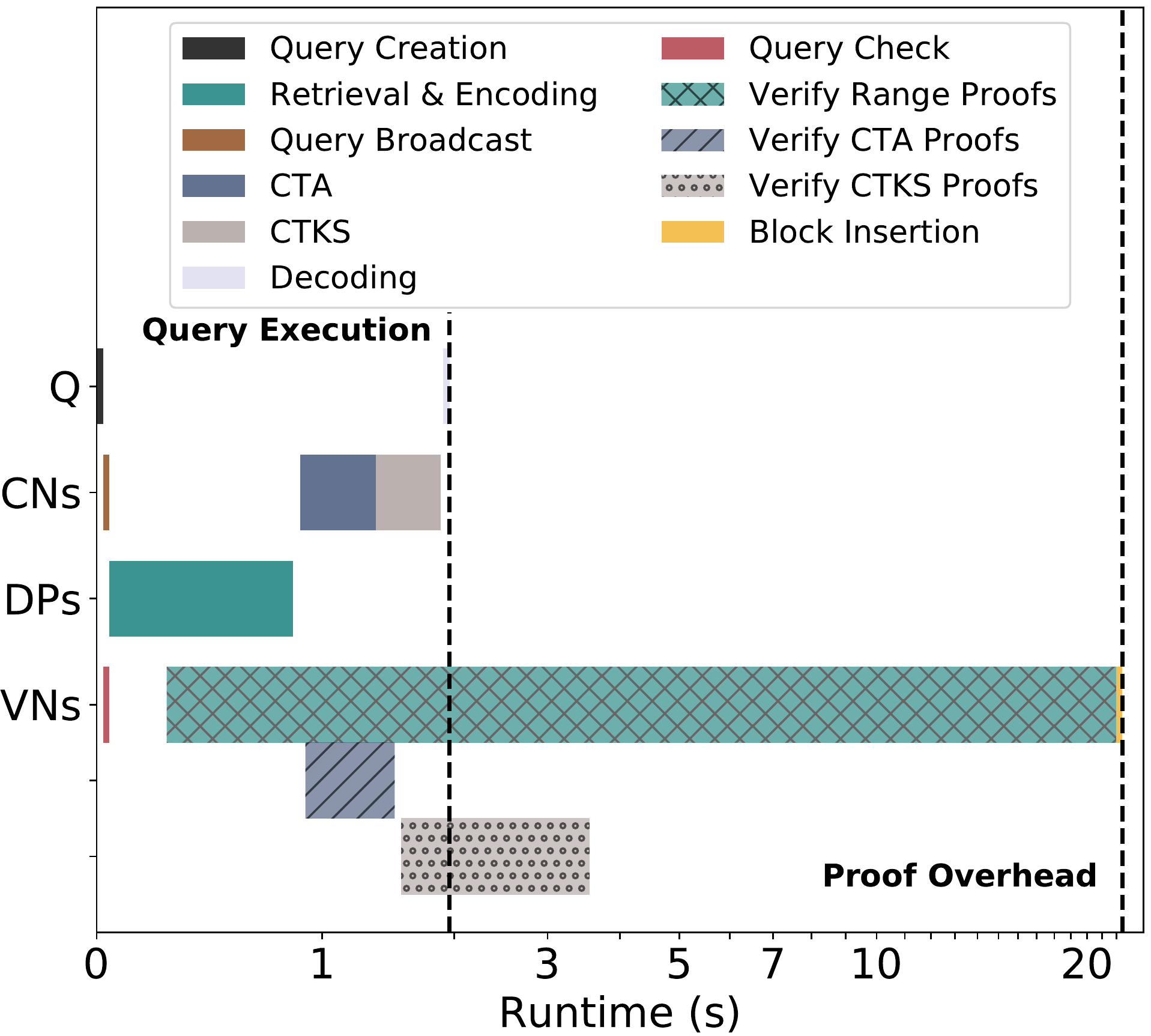}
		\vspace{-2em}
		\caption{\david{\texttt{Log. reg. training}:\\12 features, 600,000 records, max. iter.: 100.}}
		\label{fig:graph1}
	\end{subfigure}
	\begin{subfigure}[t]{0.306\textwidth}
		\centering
		\centering
		\includegraphics[width=1.0\columnwidth]{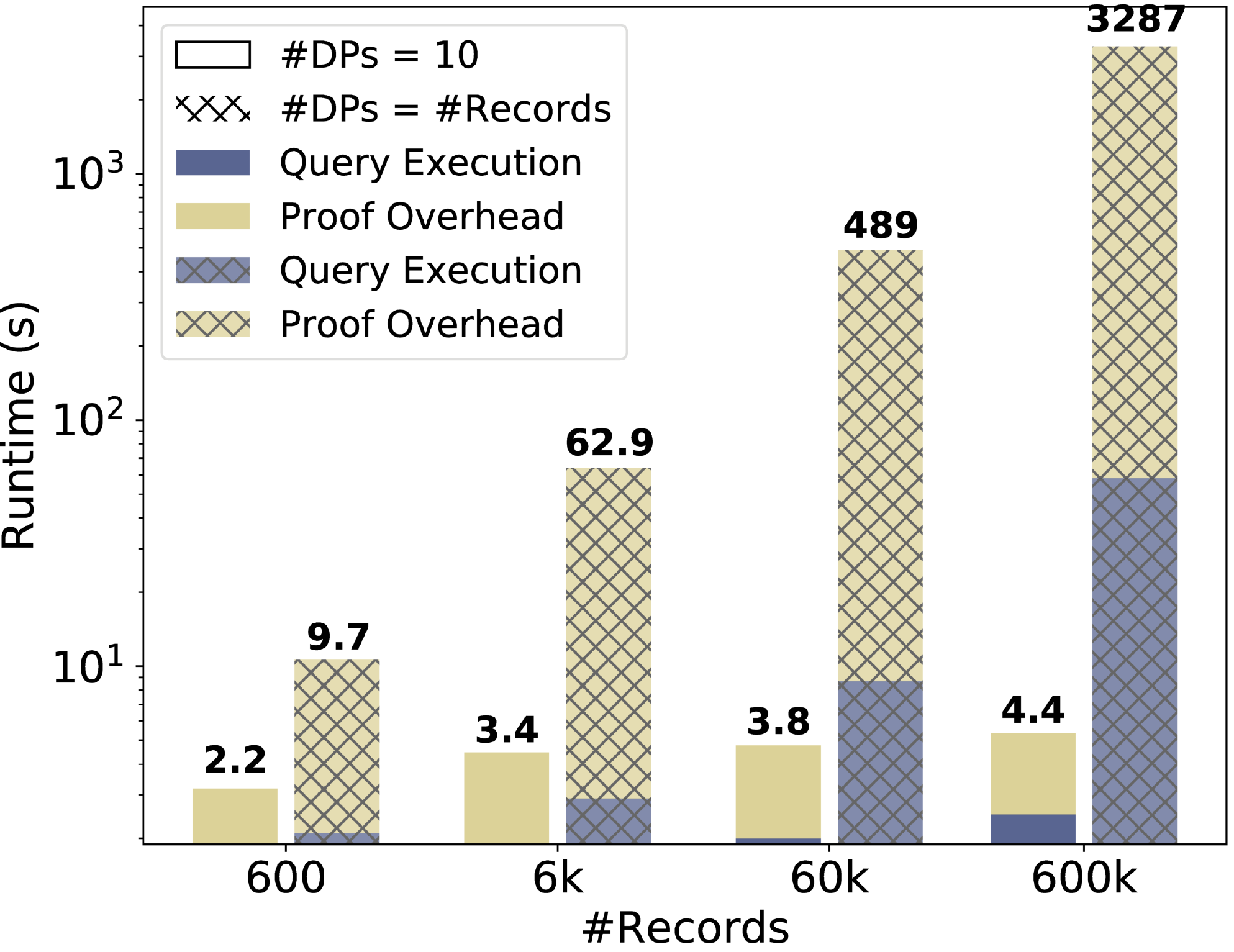}
		\vspace{-2em}
		\caption{\david{\texttt{Variance}: increasing nbr. of records.}}
		\label{fig:graph31}
	\end{subfigure}
	\vspace{-1.5em}
	\begin{subfigure}[t]{0.3\textwidth}
		\centering
		\includegraphics[width=1.0\columnwidth]{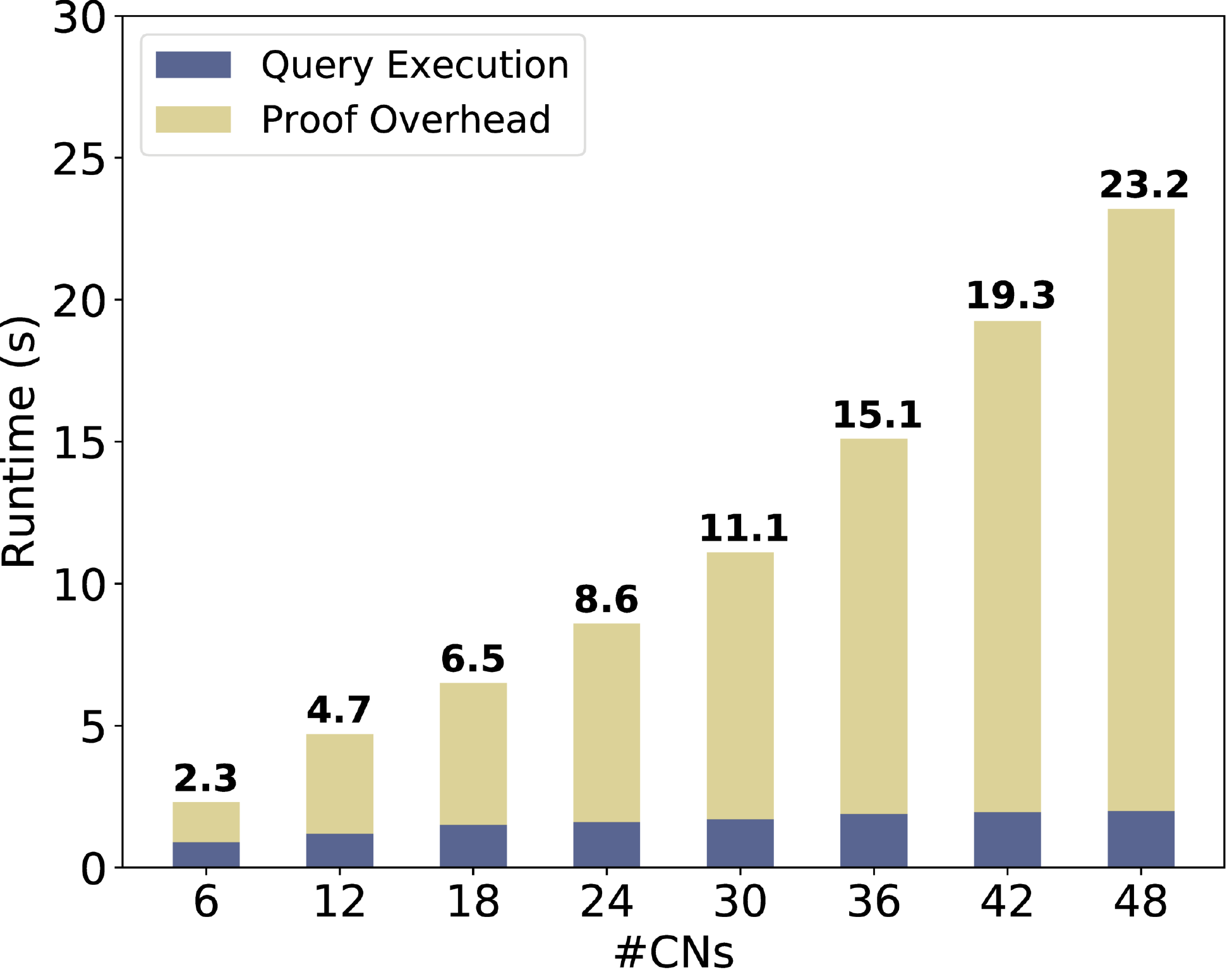}
		\vspace{-2em}
		\caption{\texttt{Variance}: increasing nbr. of $CNs$ and $DPs$ with 10 $DPs$ per $CN$.\\ }
		\label{fig:graph22}
	\end{subfigure}
	\begin{subfigure}[t]{0.3\textwidth}
		\centering
		\includegraphics[width=1.0\columnwidth]{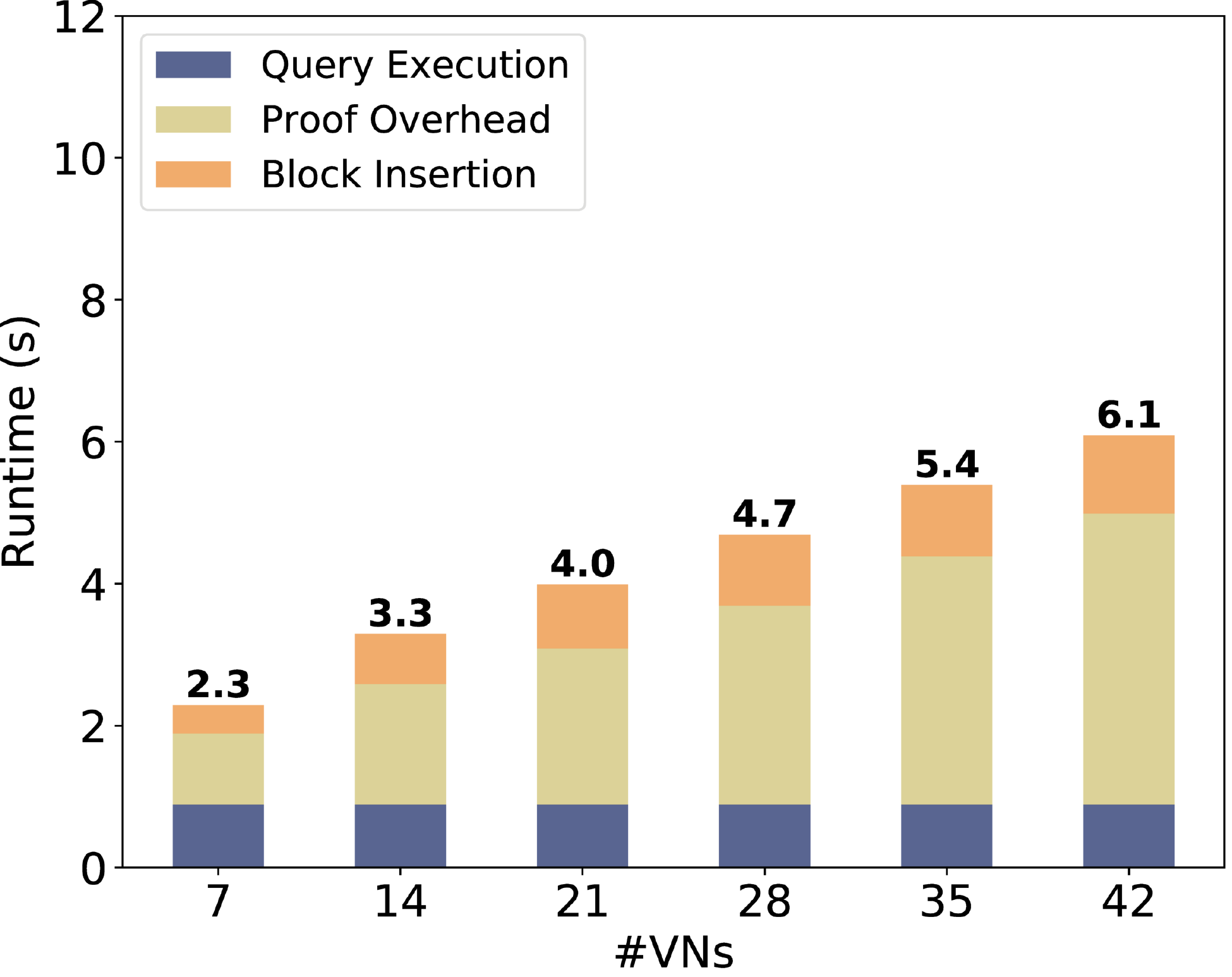}
		\vspace{-2em}
		\caption{\texttt{Variance}: increasing nbr. of $VNs$.}
		\label{fig:graph23}
	\end{subfigure}%
	\begin{subfigure}[t]{0.3\textwidth}
		\centering
		\includegraphics[width=1.0\columnwidth]{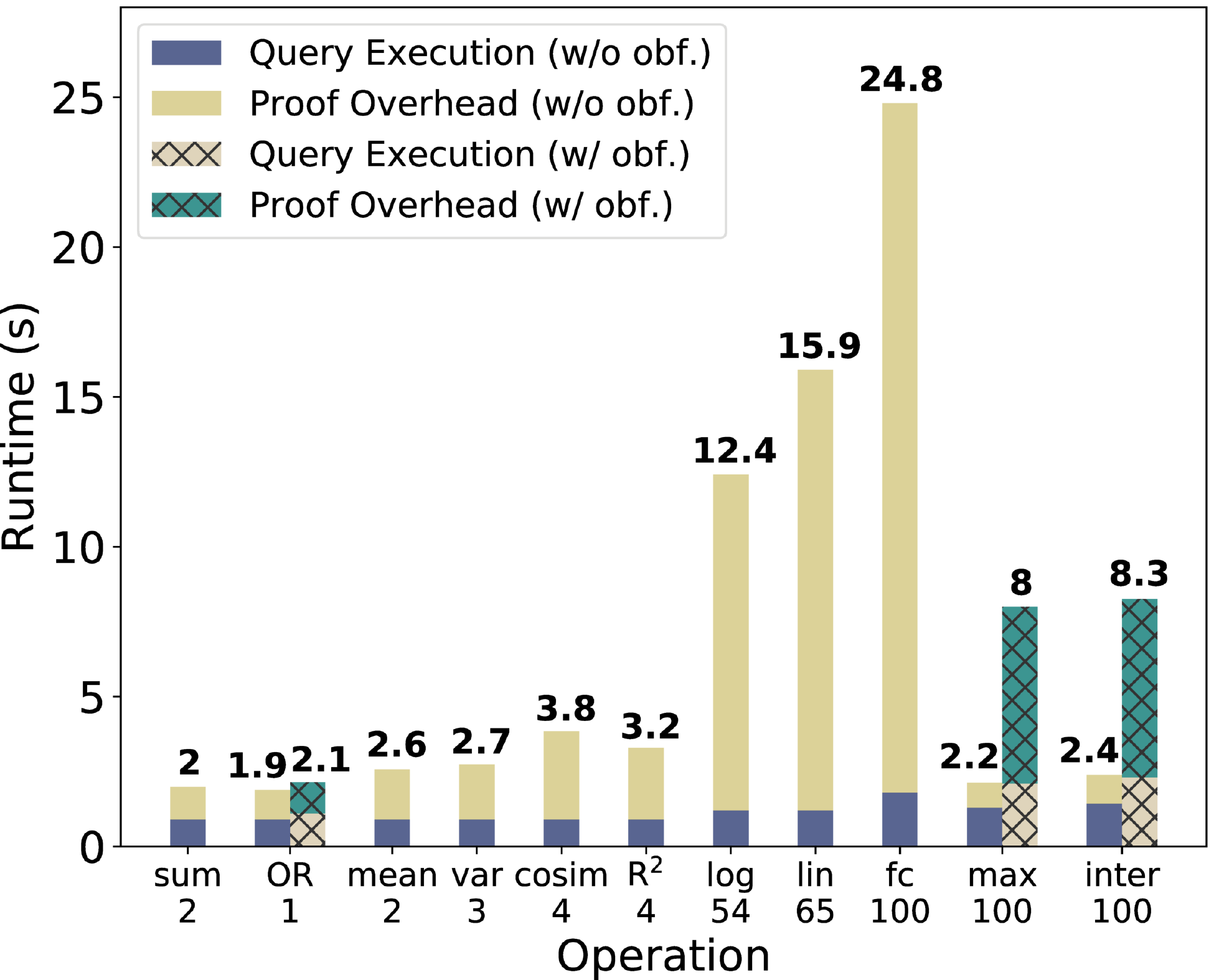}
		\vspace{-2em}
		\caption{Runtime for different operations with\\$DPs$' inputs sizes. Range $[0,2^{20}]$.}
		\label{fig:graph21}
	\end{subfigure}%
	%\vspace{-0.2em}
	\begin{subfigure}[t]{0.294\textwidth}
		\centering
		\includegraphics[width=1.0\columnwidth]{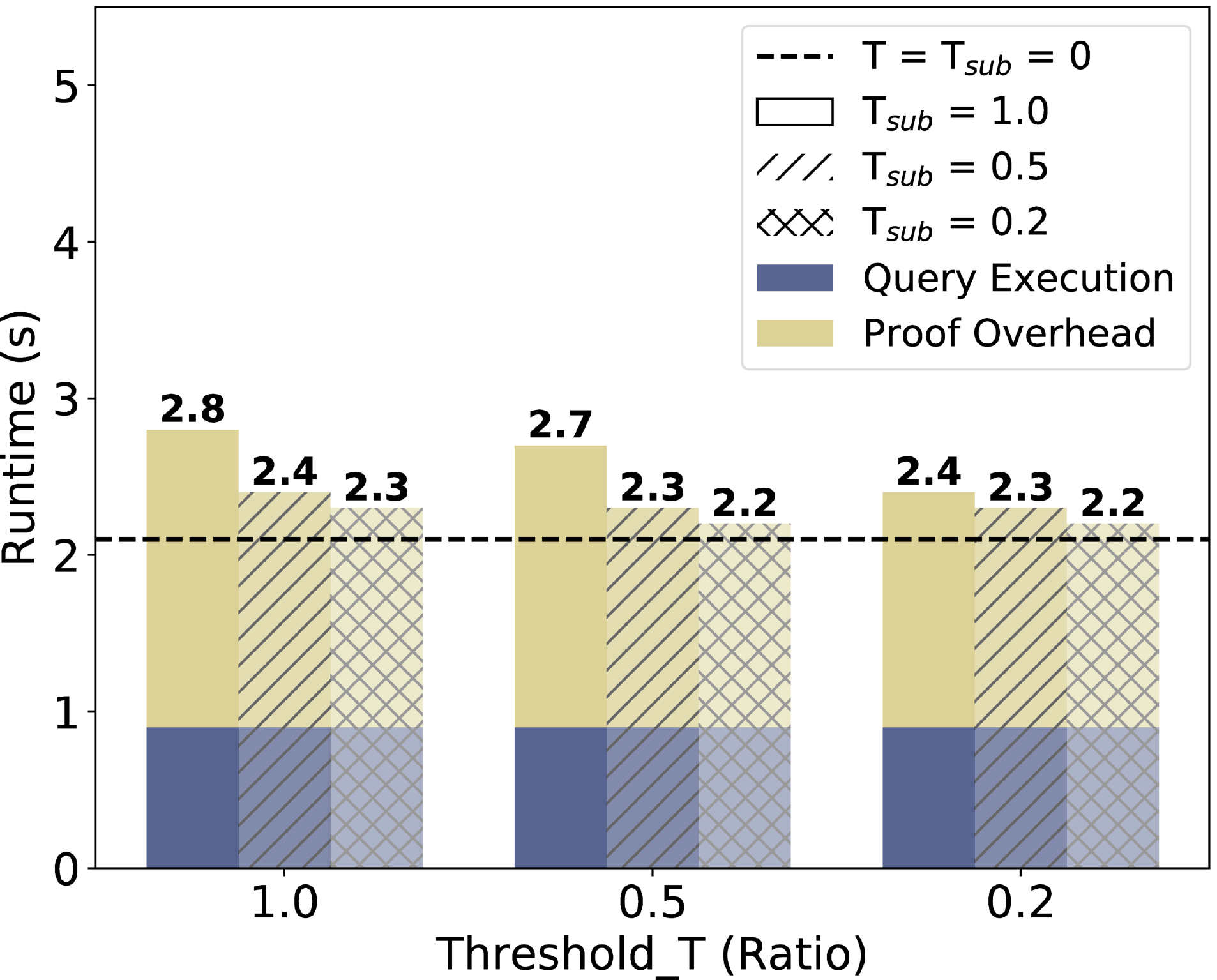}
		\vspace{-2em}
		\caption{\texttt{Variance}: proofs verif. thresholds.\\}
		\label{fig:graph32}
	\end{subfigure}
	\begin{subfigure}[t]{0.303\textwidth}
		\centering
		\includegraphics[width=1.0\columnwidth]
		{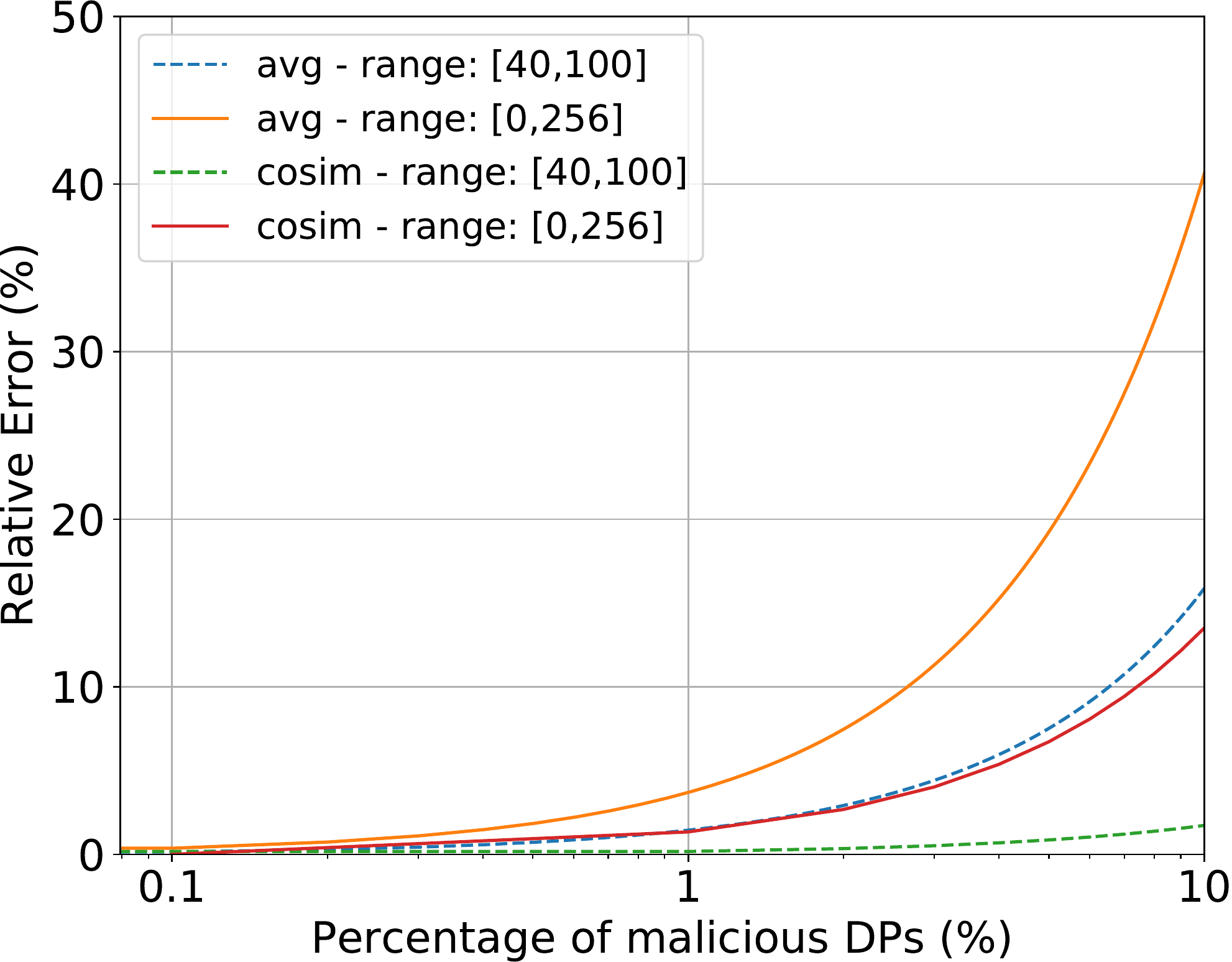}
		\vspace{-2em}
		\caption{\david{\texttt{Average} and \texttt{cosim}: influence of malicious DPs.}}
		\label{fig:maliciousDPs}
	\end{subfigure}
	\vspace{-1.0em}
	\begin{subfigure}[t]{0.3\textwidth}
		\centering
		\includegraphics[width=1.0\columnwidth]{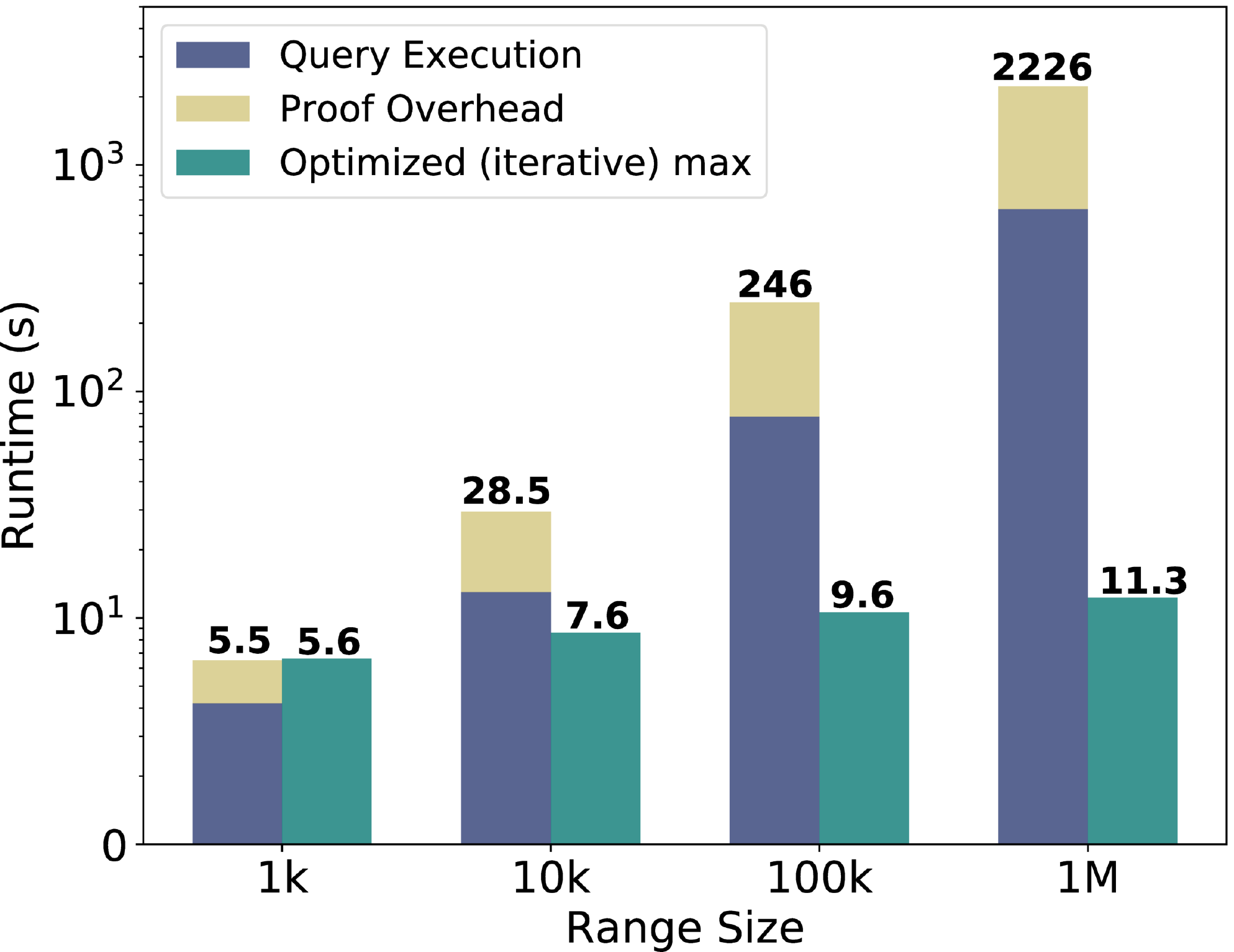}
		\vspace{-2em}
		\caption{\texttt{max} (iterative): increasing range size.}
		\label{fig:graph33}
	\end{subfigure}
	\begin{subfigure}[t]{0.294\textwidth}
		\centering
		\includegraphics[width=1.0\columnwidth]
		{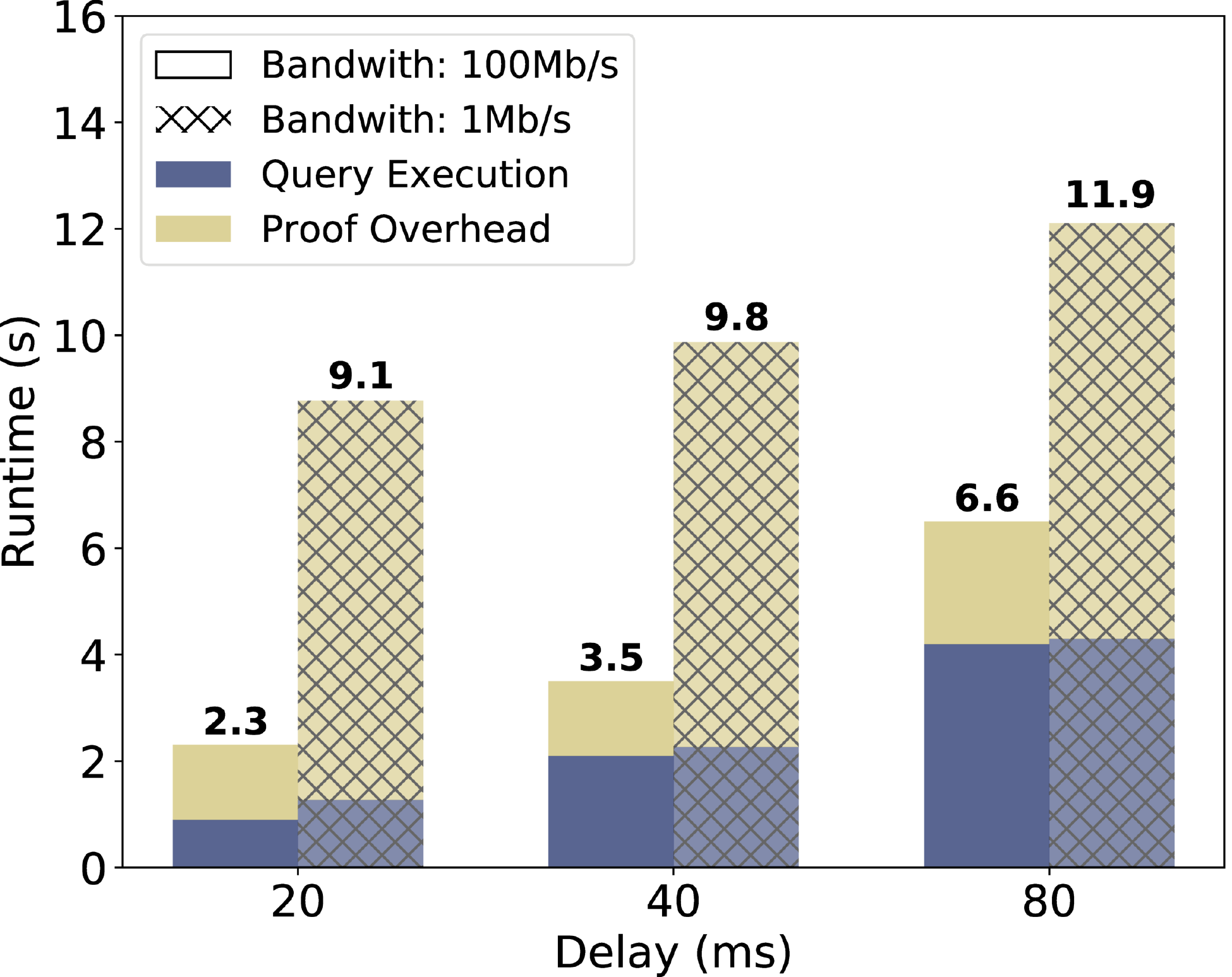}
		\vspace{-2em}
		\caption{\texttt{Variance}: runtime w.r.t. network\\ bandwidth and delay.}
		\label{fig:band}
	\end{subfigure}
	\vspace{-0.6em}
	\caption{\sys Evaluation. In Figure \ref{fig:graph33}, the optimized \textit{max} aggregates the runtime for all necessary query executions and the respective proof overheads.}
	\vspace{-1.5em}
\end{figure*}

\vspace{0.2em}
\subsubsection{\bf Drynx Evaluation}\label{drynx_evaluation}
\textbf{Parallel Execution.} Figure \ref{fig:graph1} shows the runtime for training a \textit{logistic regression} model. We use a randomly-generated dataset of \david{12 floating-point features and 600,000 records split among 12 $DPs$}.  We remark that the operations are verified in parallel to the query execution; this parallelization enables $Q$ to obtain the query results as soon as it is computed (denoted by query execution dashed line). At the end of the verification process, an auditor can check the query by verifying the signature and the \textit{query-proofs map} of the corresponding block in the \textit{proofs blockchain}, which in this case takes 0.4 seconds. The blocks' sizes are small as they only contain the query and the corresponding \textit{query-proofs map}; in this example one block is 56kB.

 \textbf{Scaling.} \david{We show how \sys's execution time evolves with an increasing number of data records (Figure \ref{fig:graph31}), $CNs$ and $DPs$ (Figure \ref{fig:graph22}) and $VNs$ (Figure \ref{fig:graph23}). Inspired by $HDS$ and $PDS$, we simulate the computation of the heart-rate \textit{variance} (values between $[0,256)$) over a set of distributed patients. In Figure \ref{fig:graph31}, we observe that \sys scales better with (a) the number of records per $DP$ (and fixed number of $DPs$) than  (b) with the number of $DPs$; case (a) represents $HDS$, where a $DP$ is an hospital with a database of multiple patients, whereas case (b) represents $PDS$, where each patient is a $DP$ ($\#DPs$ $=$ $\#records$). This is because (a) enables the $DPs$ to locally pre-aggregate their data, thus reducing the amount of proofs and computations. For Figures \ref{fig:graph22} and \ref{fig:graph23} and for the remaining part of the evaluation, we set the number of $DPs$ to 10 per $CN$. In $HDS$, this could correspond to a use case in which some $DPs$ are hospitals and the others are independent doctors sharing their data. We observe that \sys's runtime increases with the number of $DPs$, $CNs$, and $VNs$. However, an increasing number of $CNs$ and $VNs$ also means a higher security level, as the trust is distributed among more entities.}  % In Figures \ref{fig:graph22} and \ref{fig:graph23}, the $DPs$ are patients ($PDS$); and they are either patients ($PDS$) or hospitals ($HDS$) in Figure \ref{fig:graph31}.
 
\textbf{Operations.} Figure \ref{fig:graph21} shows \sys's runtime for all the operations with a large integer range of $[0,2^{20}]$ for each of the $DPs$' inputs (the size of the $DPs$' inputs is shown below each operation). We observe that for all operations, the query execution time is always below 1.5 seconds; and the overhead incurred by the proofs verification increases with the size of the $DPs$' inputs. This is expected, as the larger the DPs' inputs become, the more ciphertexts there are for the system to process, and more proofs there are to verify. We also observe that bit-wise operations take more time when the $DPs$ opt to send a bit value that is then obfuscated (using the $CTO$ protocol).
 
 \textbf{Verification Thresholds.} In Figure \ref{fig:graph32}, we show how the different thresholds on the proofs verification affect \sys's performance with a \textit{variance} query. It can be seen that sending the proofs (communication time is denoted by a dashed line) is the most time consuming part, and that reducing the thresholds reduces the verification time. For example, by having $T=1$ and $T_{sub}=0.2$, we effectively reduce the verification workload by a factor close to 0.8, and a \textit{sub-proof} is still verified by $f_h=5$ of the $VNs$ with a high probability ($83.48\%$).
 
 \textbf{Malicious DPs.} By enforcing $DPs$' values to be within a specific range, \sys limits the influence of malicious $DPs$ on the computed statistic. We illustrate this in a simple and realistic example (using $PDH$ from Section \ref{sub_ch:running_ex}) by computing the \texttt{average} heart rate over a dataset of 8922 hypertensive patients \cite{lorgis2009heart}. The real heart-rate values are limited to be between 40 bpm (beats per minute) and 100 bpm and, as presented by Lorgis et al. \cite{lorgis2009heart}, the average value obtained among honest $DPs$ is $a_h=$70 bpm with a 95\% confidence interval of $\pm 6$ bpm. Each patient ($DP_i$) must send $(\mathbf{V_i}, c_i)=([heart\_rate],count)$ (Definition \ref{def:encoding}), in which $heart\_rate$ has to be in $[40,100]$ and $count$ in $[0,1]$. In order to maximize the result's distortion, a malicious $DP$ can send an extreme value, which is within the range bounds. We assume that all malicious $DPs$ collude and send the same value $heart\_rate=e$, and that the computed average is given by $a_{m} = (h\cdot a_h + e \cdot d)/(h+c)$, where $h$ and $d$ are the numbers of honest and dishonest $DPs$, and $c$ is the sum of $c_i$ sent by malicious $DPs$. The relative error is $|1-(a_{m}/{a_h})|$. We remark that a malicious $DP$ can maximize this error with a valid input by sending $([100],0)$. In Figure \ref{fig:maliciousDPs}, we observe that with 1\% of malicious $DPs$ for the range [40,100], the highest relative error is $1.44\%$. This error corresponds to 1 bpm, still in the 95\% confidence interval. \david{We observe similar results when the cosine similarity is computed in the same settings. For this example, we also present the worst-case scenario in which the cosine similarity computed on the honest $DPs$ is 1 and the malicious $DPs$ input extreme values from the range of accepted values to reduce the similarity.} As shown in Figure \ref{fig:maliciousDPs}, these numbers highly depend on the chosen bounds. Even if many other factors influence this error (e.g., the computed operation and the distribution of the values), it shows that \sys can limit the power of malicious $DPs$. %\david{This can be extended to other operations.}
 
 %We compute again the \textit{variance} and assume that all honest $DPs$ have the same heart rate of $x = 92 bpm$ and thus send `$\mathbf{V_i} = [92, 92^2],1$' as defined in the \texttt{variance} \enco. In this case, the correct variance is 0 and, in the worst-case scenario, a malicious $DP_m$ could send `$\mathbf{V_m} = [0, 256^2],1$' in order to maximize its effect on the result. We observe that 100 $DPs$ (1\% of $DPs$) can distort the result by $1.12\%$. For this example, we consider $distortion$ to be the fraction (\%) of the maximum variance by which malicious $DPs$ managed to skew the result. We have $distortion = ((m-m^2)x^2+mb_u^2-m^2b_l^2)/(b_u^2-b_l^2)$, where $m$ is the fraction of malicious $DPs$, and $b_u$, $b_l$ are the upper and lower bounds of the range. Even though these numbers highly depend on which operation/\enco we use and on the application, it shows that \sys can limit the power of malicious $DPs$.
 \textbf{Iterative Queries.} Figure \ref{fig:graph33} depicts how \sys's runtime can be reduced by using multiple queries to execute a \texttt{min/max} operation in a binary-search style. This represents a tradeoff between privacy and performance, as each iterative query is sent in clear, leaking the interval where the min/max value is. We assume that $Q$ sets the entropy limit $EL=100$, in other words, another entity in the system can learn that the \texttt{min/max} is in an interval of at least 100 values. The precise value is kept private. We observe that the execution time is not improved when the range is small, but is greatly reduced when the range grows, reaching an execution time reduction of almost $96\%$ at a range size of 100,000.
 
\textbf{Communication.} Figure \ref{fig:band} depicts \sys's runtime evolution with respect to both the communication delay and bandwidth capacity with a heart rate \textit{variance} query. We remark that when the latter is reduced by a factor 100, the runtime increases by a factor 2 or 3. This shows that our system is more sensitive to communication delay than bandwidth capacity.
 
\begin{table}
\centering
    \ifcomment    
    \includegraphics[width=1\columnwidth]{figures/bandBlue}
    \else
	\includegraphics[width=0.7\columnwidth]{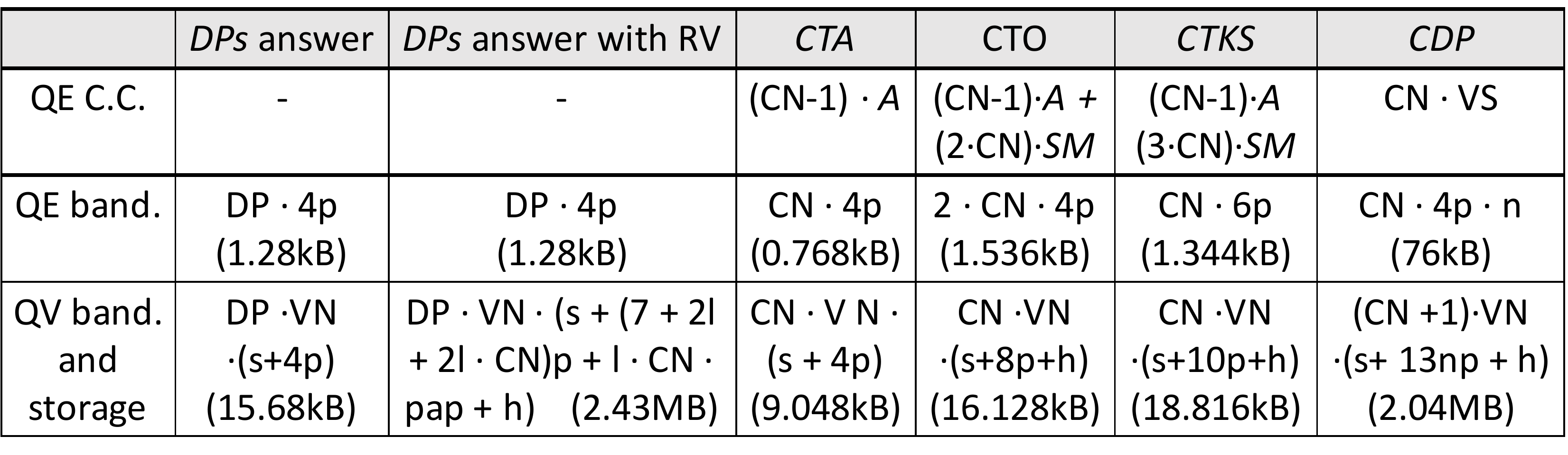}
	\fi
	%\vspace{-2.0em}
	\caption{\small{\david{Bandwidth and Storage costs. DP, VN, CN = nbr. of entities; RV = Range Valid.; QE=query exec.; C.C.=comput. complexity; QV=query verif.; \textit{A}=ciphertext addition; \textit{SM}= scalar multi.; \textit{VS}= verif. shuffle}}}
	\label{tab_band}
	\vspace{-2.0em}
\end{table} 

\begin{figure*}[ht!]
	\centering
	\tiny
	\begin{subfigure}[t]{0.6\textwidth}
		\centering
		\includegraphics[width=1\columnwidth]
		{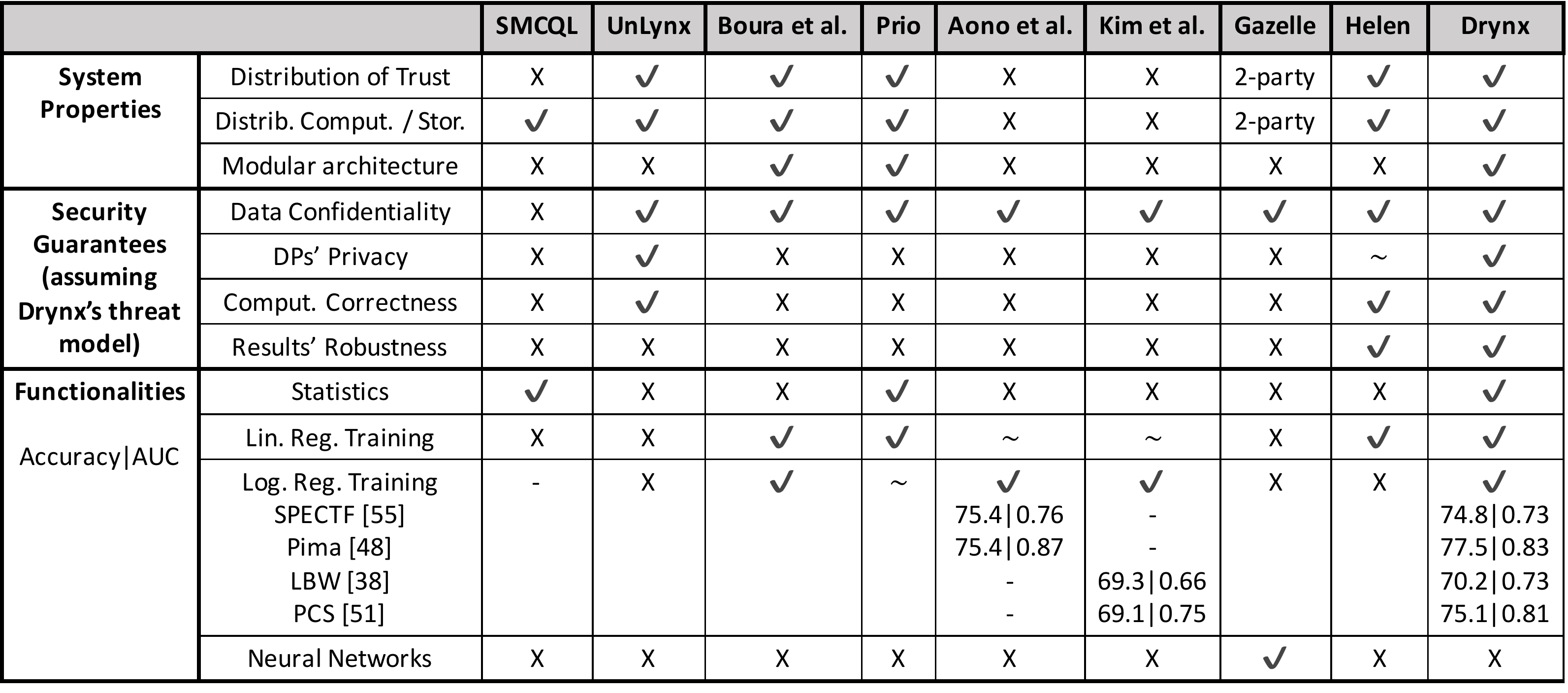}
		%\vspace{-1.5em}
		\caption{Solutions Comparison. For log. reg., we split the datasets \cite{spectf, pima,pcs, lbw}  among 10 $DPs$ before standardization, scale factor $=10^2$ for fixed-point represent., learning rate 0.1; 80\% train., 20\% test. \nocite{spectf, pima,pcs, lbw}}
		\label{tab:compare}
	\end{subfigure}
\hspace{2em}
	\begin{subfigure}[t]{0.32\textwidth}
		\centering
		\includegraphics[width=1.0\columnwidth]
		{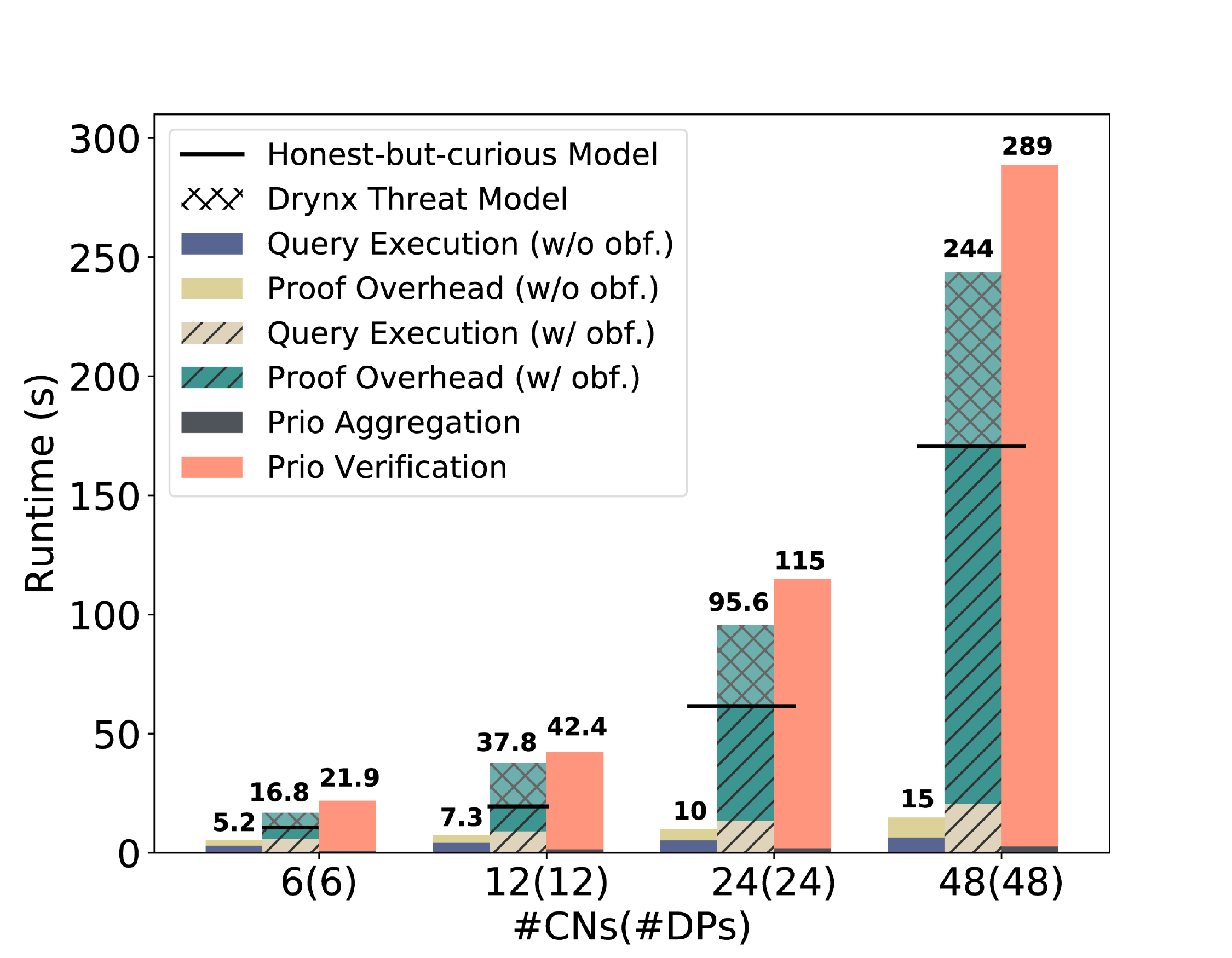}
		\vspace{-1.5em}
		\caption{Comparison with Prio for a \textit{min} query.}
		\label{fig:graph4}
	\end{subfigure}
	\vspace{-1.5em}
	\caption{\sys's comparisons.}
	%\vspace{-4.0em}
\end{figure*}

\textbf{Bandwidth.}  In Table \ref{tab_band}, we present the computation and bandwidth complexities for 1 ciphertext (i.e., 2 points (2p) on the Elliptic Curve, 2p = 64 bytes) per $DP$. We use $DP$, $VN$, and $CN$ as the numbers of corresponding entities in the system. $s$ is the size of the Schnorr signature \cite{fiat1986prove} ($s=96$ bytes), $h$ is the hash size ($h=32$ bytes), $l$ comes from the range $[0,u^l)$ for the range proofs ($u^l=16^2, l=2$), $pap$ is a pairing point's size ($pap=384$ bytes) and $n$ is the number of values that are used in the $CDP$ ($n=100$). \david{We do not include the computational complexity for the local computations executed by the $DPs$ and $CNs$. We refer to Neff's work \cite{neffverifiable} for the complexity of the verifiable shuffle ($VS$). We observe that when the number of $CNs$ and $VNs$ increases, the computational, bandwidth and storage costs increase for all the steps. As having more $CNs$ or $VNs$ improves the security and the distribution of the workload in the system, it creates a tradeoff between security, efficiency, and scalability.} %We also notice that the overhead of the input-range validation proofs and $$ are non-negligible. Nevertheless, they remain acceptable as the range proofs for one $DP$ use, in our example, up to 5.8kB on one $VN$.

\subsubsection{\bf Comparison with Existing Works}\label{drynx_comparison}
We supplement the related work's overview, described in Section \ref{sec:related}, by presenting here a qualitative and quantitative comparison with multiple systems that are \sys's closest related works. We compare \sys against SMCQL \cite{bater2017smcql}, UnLynx \cite{froelicher2017unlynx}, Prio \cite{corrigan2017prio}, Boura et al. \cite{bourahigh}, Aono et al. \cite{aono2016scalable}, Kim et al. \cite{kim2018secure} and Gazelle \cite{juvekar2018gazelle}. In Table \ref{tab:compare}, we show that \sys provides several functionalities in a strong threat model and achieves results that can rival with other secure and dedicated approaches, notably in the training of logistic regression models as depicted in Figure \ref{tab:compare}. \sys performs as well or better than its two closest related works, UnLynx and Prio, and provides better security guarantees.

We observe that solutions based exclusively on secret sharing and garbled circuits, namely SMCQL \cite{bater2017smcql}, Prio \cite{corrigan2017prio} and Boura et al. \cite{bourahigh}, offer multiple or advanced functionalities but fail to provide proofs of correct executions. Systems solely based on homomorphic encryption (HE), namely UnLynx \cite{froelicher2017unlynx}, Aono et al. \cite{aono2016scalable}, Helen \cite{popa2019helen} and Kim et al. \cite{kim2018secure}, are limited in the functionalities they offer. Furthermore, Aono et al. \cite{aono2016scalable} and Kim et al. \cite{kim2018secure} rely on data centralization. Gazelle \cite{juvekar2018gazelle} combines HE and garbled circuits and enables complex evaluations of neural networks, but does not protect $DPs$' privacy or provide computation correctness. Contrarily, \sys enables multiple operations while distributing trust, computations, and data storage, and it provides strict security guarantees in a stronger adversarial model. 

We quantitatively compare \sys to Unlynx \cite{froelicher2017unlynx} and Prio \cite{corrigan2017prio}, which are, to the best of our knowledge, the closest prior works. \sys's query execution time for the \texttt{sum} is faster than UnLynx, as we improved the $CTKS$ protocol by enabling its execution in a tree fashion, thus reducing its execution complexity from $O(\#CN)$ to $O(log(\#CN))$. Unlike UnLynx, \sys enables the verification of $DPs$' value ranges, which, for the computation of a \texttt{sum}, adds an overhead of only 0.6 seconds (out of a total time of 2 seconds, as depicted in Figure \ref{fig:graph21}). However, \sys enables a faster scalable verification of proofs by an auditor. After the proofs are verified and the results stored in the \emph{proof blockchain}, an auditor can simply request and verify the corresponding block, which in this case takes approximately 0.4s. In Unlynx, an auditor has to request the proofs from each entity and verify them by itself, which takes 1.4s.

Prio \cite{corrigan2017prio} relies on secret-shared non-interactive proofs that are created by the $DPs$ to prove the correctness of their inputs to the system and that are collectively verified by the $CNs$. Even though both systems have similar functionalities, Prio provides input-range verification and computation correctness only when all the $CNs$ are honest-but-curious. We adapted the Gorrigan-Gibbs prototype implementation \cite{prioCode} of Prio to a similar deployment environment as \sys so that both use the same communication settings, thus enabling a fair comparison. In Figure \ref{fig:graph4}, we compare Prio's runtime in an illustrative example by using the \texttt{min} operation on the range $[0,1000)$ with increasing number of $CNs$ and $DPs$, against multiple settings of \sys.  This figure shows that \sys significantly outperforms Prio when computing \texttt{min} without using obfuscation ($CTO$) hence accepts a small probability of error (${1}/{(\#\mathcal{G})}$) and avoids the need for range proofs. If we use obfuscation, \sys scales similarly as Prio, but it must be noted that Drynx performs its operations in a stronger threat model. When used in Prio's threat model (delimited by a black line), \sys is about two times faster. This is because each range proof can be sent and verified by a single $VN$ as all $VNs$ are considered honest-but-curious under Prio's threat model.

%% file: conclusion.tex
\vspace{-0.5em}
\section{Conclusion}\label{sec:conclusion}
We have proposed \sys, a novel system that enables a querier to compute statistics and train machine-learning models on distributed datasets in a strong adversarial model where no entity is individually trusted. \sys{} provides query-execution auditability and ensures the end-to-end confidentiality of the data. It protects the privacy of the data providers and relies on an immutable and distributed ledger to provide efficient correctness verification and proofs storage. \sys is highly modular, offering configurable tradeoffs between security, privacy, and efficiency. Finally, \sys enables privacy-preserving computations of widely-used statistics on sensitive and distributed data, thus offering features that are absolutely needed in crucial areas such as user-behavior analysis or research for personalized medicine. 

%% file: appendix.tex
%\vspace{-1em}
\newpage
\appendices
%\begin{appendices}
\section{Table of Symbols}\label{sym_table}
\vspace{-1em}
\begin{table}[h!]
	\centering
	\small
	\begin{tabular}{| c | c |}
		\hline
		\textbf{Symbol} &  \textbf{Description} \\ \hline
		$HDS$, $PDS$ & Hospitals \& Patients Data Sharing\\   \hline
		$\mathcal{G}$, $B$, $p$ & Elliptic curve; base point on $\mathcal{G}$, prime	 \\ \hline
		E$_\Omega$($m$) = $\scriptsize (C_1, C_2)$  & ElG encrypt. of $m$ under key $\Omega$, \\ 
		$ \scriptsize=(rB,\ mB+r\Omega)$ &   nonce $r$ \\ \hline
		$K$  & $CNs$ pub. coll. key  \\   \hline
		($k_i$, $K_i$)  & $CNs$ $CN_i$ priv., pub. key  \\   \hline
		$A$, $A_1$, $A_2$, $Y_i$, $y_i$ & ZKPs pub. (uppercase), discrete log.	 \\ \hline
		$Q$, $DP$, $N$ & Querier, Data Provider, $\#DP$ \\   \hline
		$CN$, $VN$ & Computing \& Verifying Node \\   \hline
		$f_h$ & Threshold of honest $VNs$ \\ \hline
		$\pi$, $\rho$, $\bar{r_i}$ & linear combi., \enco, records	 \\ \hline
		$\mathbf{V_i}=[v_{i,1}, ..., v_{i,l}]$, $c_i$ & vector, count	 \\ \hline
		$CTA$, $CTO$, $CTKS$ & Coll. Tree Aggr., Obfusc., Key Switch. \\   \hline
		$w_{i,1}, w_{i,2}$ & $CN_i$'s contribution in $CTKS$ \\ \hline
		$\alpha_i$, $s_i$ & $CN_i$ secret random nonce \\ \hline
		$CDP$, ($\epsilon$, $\delta$, $\theta$) & Coll. Diff. Privacy \& params.\\ \hline
		%$EL$, ($\epsilon$, $\delta$, $\theta$), $n$ & Privacy params. \\ \hline
		$[b_l,b_u]$, $[0,u^l)$ & Range, default range  \\ \hline %$b_l$, $b_u$, $u$, $l$ integers  \\ \hline
		$x_i$, ($A_{i,j}$, $Z_i$, $H$, $V_{i,j}$, $a_{i,j}$) & Range proof priv., pub. values   \\ \hline
		$T$, $T_{sub}$ & Proofs and sub-proofs verif. thresh. \\ \hline
		$N_{da}, N_{i}, D$ & Tot. \& $DP_i$  $\#$records, dataset dim. \\ \hline
		%$b_i, x_i$, $D$ & $DP_i$ bit, int, input size \\ \hline
		%$d$ & Dataset dimension \\ \hline
		%$[\tilde{n}_{\tilde{1}},..,\tilde{n}_{\tilde{l}}], [\hat{n}_{\tilde{1}},..,\hat{n}_{\tilde{l}}]$,$LD$ & Noise values, Laplace Distribution \\ \hline
		$p_{ver}$, $p_{ver_{sub}}$ & proof, sub-proof\\ \hline
		$P_{f_h}$ & prob. of $f_h$ $VNs$ verif. \\ \hline
		%$pap, s$ & pairing point, Schnorr sign. \\ \hline
		%$h$ & hash \\ \hline
	\end{tabular}
	\vspace{-0.7em}
	\caption{Table of Recurrent Symbols.}
	\label{tab:symbols}
	%\vspace{-1.5em}
\end{table}
%\vspace{-1em}
\section{Error Probability}\label{error_proba}
%\vspace{-0.5em}
%We derive here the expression for the probability of error $P_n$ described in Section \ref{sec:encoding}.
In Section \ref{sec:encoding}, we notice that the result of bit-wise operations, when $DPs$ are requested to answer with random values $R_is$, can be erroneous with a probability smaller than ${1}/{(\#G-1)}$. We demonstrate here this result and provide an expression for the probability of error $P_n$ where $n$ is the number of $DPs$.
\begin{align*}
P_n & \scriptstyle = P(\scriptstyle\sum\limits_{i=1}^{n}  R_i = 0)  = \scriptstyle\sum\limits_{a=0}^{\#G-1} P(\scriptstyle\sum\limits_{i=1}^{n}  R_i = 0 \mid \scriptstyle\sum\limits_{i=1}^{n-1}  R_i = a) \cdot P(\scriptstyle\sum\limits_{i=1}^{n-1}  R_i = a) 
\\ &\scriptstyle  =  P(\scriptstyle\sum\limits_{i=1}^{n}  R_i = 0 \mid \scriptstyle\sum\limits_{i=1}^{n-1}  R_i = 0) \cdot P(\scriptstyle\sum\limits_{i=1}^{n-1}  R_i = 0) 
\\ &\scriptstyle  + \scriptstyle\sum\limits_{a=1}^{\#G-1} P(\scriptstyle\sum\limits_{i=1}^{n}  R_i = 0 \mid \scriptstyle\sum\limits_{i=1}^{n-1}  R_i = a) \cdot P(\scriptstyle\sum\limits_{i=1}^{n-1}  R_i = a) 
\\ &\scriptstyle  = \scriptstyle\sum\limits_{a=1}^{\#G-1} P(\scriptstyle\sum\limits_{i=1}^{n}  R_i = 0 \mid \scriptstyle\sum\limits_{i=1}^{n-1}  R_i = a) \cdot P(\scriptstyle\sum\limits_{i=1}^{n-1}  R_i = a) 
\\ &\scriptstyle  = P(R_n = -a) \cdot \scriptstyle\sum\limits_{a=1}^{\#G-1}  P(\scriptstyle\sum\limits_{i=1}^{n-1}  R_i = a) 
\\ &\scriptstyle  = \scriptstyle\frac{1}{\#G-1} \cdot \scriptstyle\sum\limits_{a=1}^{\#G-1}  P(\scriptstyle\sum\limits_{i=1}^{n-1}  R_i = a)  = \scriptstyle\frac{1}{\#G-1} \cdot (1-P_{n-1}).
\end{align*}
We have $\scriptstyle P_n = \scriptstyle\frac{1}{\#G-1} \cdot (1-P_{n-1}) \leq  \scriptstyle\frac{1}{\#G-1}$
and $\scriptstyle P_n = \scriptstyle\sum\limits_{i=2}^{n}  (-1)^i\cdot (\scriptstyle\frac{1}{\#G-1})^{i-1}$.
%\vspace{-1em}

%\end{appendices}